  \documentclass{emulateapj}

\usepackage{graphicx}
\usepackage{epstopdf}

\newcommand{\gsim}{ {}^>_\sim}

\newcommand {\Lya}    {Ly$\alpha$}   
\newcommand {\Lyb}    {Ly$\beta$}    

\newcommand {\HI}     {\ion{H}{1}}      
\newcommand {\HeI}    {\ion{He}{1}}   
\newcommand {\HeII}   {\ion{He}{2}}   

\newcommand {\OI}      {\ion{O}{1}}       
\newcommand {\OII}     {\ion{O}{2}}       
\newcommand {\OIII}    {\ion{O}{3}}       
\newcommand {\OIV}     {\ion{O}{4}}       
\newcommand {\OV}      {\ion{O}{5}}        
\newcommand {\OVI}     {\ion{O}{6}}        

\newcommand {\NI}      {\ion{N}{1}}

\newcommand {\NeV}    {\ion{Ne}{5}}   
\newcommand {\NeVIII} {\ion{Ne}{8}}   

\newcommand {\NHI}    {$N_{\rm HI}$}

\newcommand {\kms}    {km~s$^{-1}$}

\newcommand {\FUSE}   {{\it FUSE}} 
\newcommand {\HST}    {{\it HST}}
\newcommand{\IUE}     {{\it IUE}}  
\newcommand{\Galex} {{\it GALEX}}

\newcommand {\etal}   {et~al.} 

\begin{document}

\title{{\it HST}-COS Observations of AGN. II.  Extended Survey of Ultraviolet Composite Spectra from 
159 Active Galactic Nuclei\footnote{Based on observations made with the NASA/ESA {\it Hubble Space Telescope}, 
obtained from the data archive at the Space Telescope Science Institute. STScI is operated by the Association of 
Universities for  Research in Astronomy, Inc. under NASA contract NAS5-26555.}  }  

\author{Matthew L. Stevans\altaffilmark{1}, J. Michael Shull\altaffilmark{2}, Charles W. Danforth, and Evan M. Tilton}
\affil{CASA, Department of Astrophysical \& Planetary Sciences, \\
University of Colorado, Boulder, CO 80309, USA}

\altaffiltext{1}{Now at Astronomy Department, University of Texas, Austin, TX, 78712}  
\altaffiltext{2}{Also at Institute of Astronomy, Cambridge University, Cambridge CB3~OHA, UK} 

\email{stevans@astro.as.utexas.edu,   \\  michael.shull@colorado.edu }  


\begin{abstract} 
The ionizing fluxes from quasars and other active galactic nuclei (AGN) are critical for interpreting their 
emission-line spectra and for photoionizing and heating the intergalactic medium (IGM).   Using 
far-ultraviolet spectra from the Cosmic Origins Spectrograph (COS) on the {\it Hubble Space Telescope} 
(\HST), we directly measure the rest-frame ionizing continua and emission lines for 159 AGN at redshifts 
$0.001 < z_{\rm AGN} < 1.476$ and construct a composite spectrum from 475--1875~\AA.  We identify the
underlying AGN continuum and strong EUV emission lines from ions of oxygen, neon, and nitrogen after
masking out absorption lines from the \HI\ \Lya\ forest, 7 Lyman-limit systems ($N_{\rm HI} \geq 10^{17.2}$ 
cm$^{-2}$) and 214 partial Lyman-limit systems ($15.0 < \log N_{\rm HI}< 17.2$).  The 159 AGN exhibit a 
wide range of FUV/EUV spectral shapes, $F_{\nu} \propto \nu^{\alpha_{\nu}}$ typically with 
$-2 \leq \alpha_{\nu} \leq 0$ and no discernible continuum edges at 912~\AA\ (\HI) or 504~\AA\ (\HeI).  The 
composite rest-frame continuum shows a gradual break at $\lambda_{\rm br} \approx 1000$~\AA, with mean 
spectral index $\alpha_{\nu} = -0.83 \pm 0.09$ in the FUV (1200--2000~\AA) steepening to 
$\alpha_{\nu} = -1.41 \pm 0.15$ in the EUV (500--1000~\AA).  We discuss the implications of the UV flux 
turnovers and lack of continuum edges for the structure of accretion disks, AGN mass inflow rates, and 
luminosities relative to Eddington values.  

\end{abstract} 



\section{INTRODUCTION}

The far ultraviolet (FUV) and extreme ultraviolet (EUV) continua of active galactic nuclei (AGN) are thought to form 
in the black hole accretion disk (Krolik 1999;  Koratkar \& Blaes 1999), but their ionizing photons can influence 
physical conditions in the broad emission-line region of the AGN as well as  the surrounding interstellar and intergalactic
gas.  The metagalactic background from galaxies and AGN is also an important parameter in cosmological simulations,
as a dominant source of ionizing radiation, critical for interpreting broad emission-line spectra of AGN, intergalactic 
metal-line absorbers, and fluctuations in the ratio of the \Lya\ absorbers of \HI\ and \HeII. 

Since the deployment of the first space-borne ultraviolet (UV) spectrographs, astronomers have combined spectral
observations of AGN into composite spectra.  These composites constrain the intensity and shape of the AGN 
component of the ionizing photon background.  The most direct probe of the FUV and EUV continua in the AGN 
rest frame comes from observations taken by instruments such as the {\it International Ultraviolet Explorer} (IUE) 
and a series of UV spectrographs aboard the {\it Hubble Space Telescope} (\HST):  the Goddard High Resolution 
Spectrograph (GHRS), the Faint Object Spectrograph (FOS), the Space Telescope Imaging Spectrograph (STIS), 
and the Cosmic Origins Spectrograph (COS).   Ultraviolet spectra were also obtained by the {\it Far Ultraviolet 
Spectroscopic Explorer} (\FUSE).   For AGN at modest redshifts, all of these instruments provide access to the 
rest-frame Lyman continuum (LyC, $\lambda < 912$~\AA), and at $z < 1.5$ they avoid strong contamination from 
the \Lya-forest absorbers in the spectra of high-redshift AGN.   Thus, obtaining access to high-S/N, 
moderate-resolution UV spectra is crucial for finding a reliable underlying continuum. \\

This is our second paper, following a AGN composite spectrum  presented in Paper~I (Shull \etal\ 2012) based on 
initial results from COS spectra of 22 AGN at redshifts $0 < z < 1.4$.  We analyzed their rest-frame FUV and EUV 
spectra, taken with the G130M and G160M gratings, whose resolving power $R = \lambda /\Delta \lambda \approx 18,000$ 
(17 \kms\ velocity resolution) allows us to distinguish line blanketing from narrow interstellar and intergalactic absorption
lines.   Here, in  Paper II, we enlarge our composite spectrum from 22 to 159 AGN, confirm the validity of our early results, 
and explore possible variations of the EUV spectral index with AGN type and luminosity.  Both studies were enabled by 
high-quality, moderate-resolution spectra taken with the Cosmic Origins Spectrograph installed on \HST\  during the May 
2009 servicing mission.  The COS instrument (Green \etal\ 2012) was designed explicitly for point-source spectroscopy of
faint targets, particularly quasars and other AGN used as background sources for absorption-line studies  of the intergalactic 
medium (IGM), circumgalactic medium (CGM), and interstellar medium (ISM).  Our survey is based on high-quality spectra 
of the numerous AGN used in these projects. \\

Our expanded survey of 159 AGN finds a composite spectral energy distribution (SED) with frequency index 
$\alpha_{\nu} = -1.41 \pm 0.15$ in the rest-frame EUV.  This confirms the results of Paper I, where we found  
$\alpha_{\nu} = -1.41 \pm 0.21$.  We adopt the convention in which rest-frame flux distributions are fitted 
to power laws in wavelength, $F_{\lambda} \propto \lambda^{\alpha_{\lambda}}$, and converted to 
$F_{\nu} \propto \nu^{\alpha_{\nu}}$ in frequency with $\alpha_{\nu} =  -(2 + \alpha_{\lambda})$.  We caution
that these spectral indices are {\it local} measures of the slope over a small range in wavelength, 
$\Delta \lambda / \lambda \approx 0.45$.  Because of curvature of the AGN spectral distributions, 
local slopes can be misleading, when compared to different wavelength bands and to objects at different redshift.
We return to this issue in Section 3.3 where we discuss possible correlations of indices $\alpha_{\lambda}$ with
AGN type, redshift, and luminosity and compare indices measured by both \HST/COS and \FUSE.  \\

The COS composite spectrum is somewhat harder than that in earlier \HST/FOS and STIS observations (Telfer \etal\ 
2002) who fitted the continuum (500--1200~\AA) with $\alpha_{\nu}  = -1.76\pm0.12$ for 184 QSOs at $z > 0.33$.  
Their sample of 39 radio-quiet AGN had $\alpha_{\nu}  = -1.57\pm0.17$.   Our fit differs considerably
from the \FUSE\ survey of 85 AGN at $z \leq 0.67$ (Scott \etal\  2004) who found a harder composite spectrum with 
$\alpha_{\nu} = -0.56^{+0.38}_{-0.28}$.  The different indices could arise in part from the small numbers of targets 
observed in the rest-frame EUV.  Even in the current sample, only 10 or fewer AGN observations cover the spectral 
range $450~{\rm \AA} \la \lambda \la 600$~\AA.   Another important difference in methodology is our placement of the 
EUV continuum relative to strong emission lines such as \NeVIII\ (770, 780~\AA), \NeV\ (570~\AA), \OII\ (834~\AA),
\OIII\ (833, 702~\AA), \OIV\ (788, 554~\AA), \OV\ (630~\AA), and \OVI\ (1032, 1038~\AA).  Identifying and fitting these 
emission lines requires high S/N spectra.  A complete list of lines appears in Table 4 of Paper~I.  We also use the 
higher spectral resolution of the COS (G130M and G160M) gratings to distinguish the line-blanketing by narrow 
absorption lines from the \Lya\ forest.   Increasingly important at higher redshifts, we need to identify and correct for 
absorption from Lyman-limit systems (LLS) with N$_{\rm HI}~\gsim 10^{17.2}$ cm$^{-2}$ and partial Lyman-limit 
systems (pLLS) with N$_{\rm HI} = 10^{15} - 10^{17.2}$ cm$^{-2}$.   The historical boundary at $10^{17.2}$ cm$^{-2}$ 
occurs where the photoelectric optical depth $\tau_{\rm HI} = 1$ at the 912~\AA\ Lyman edge. \\

In Paper I, our 22 AGN ranged in redshift from $z = 0.026$ to $z = 1.44$, but included only four targets at sufficient 
redshift to probe the rest-frame continuum below 550~\AA.  Our new survey contains 159 AGN out to $z = 1.476$ with 
16 targets at $z > 0.90$, sufficient to probe below 600~\AA\ by observing with G130M down to 1135 \AA.
In all AGN spectra, we identify the prominent broad emission lines and line-free portions of the spectrum and fit the 
underlying continua, excluding interstellar and intergalactic absorption lines.  In Section 2 we describe the COS 
data reduction and our new techniques for restoring the continua with a fitting method that corrects for the effects of 
absorption from the IGM and ISM.   In Section~3 we describe our results on the FUV and EUV spectral indices
in both individual and composite spectra.  Section~4 presents our conclusions and their implications.  \\

\section{OBSERVATIONS OF ULTRAVIOLET SPECTRA OF AGN}  

\subsection{Sample Description} 

Table~1 lists the relevant COS observational parameters of our 159 AGN targets, which include AGN type and
redshift ($z_{AGN}$), from NED\footnote{NASA/IPAC Extragalactic Database (NED) is operated by the Jet Propulsion 
Laboratory, California Institute of Technology, under contract with the National Aeronautics and Space Administration, 
{\tt \url{http://nedwww.ipac.caltech.edu.}} }, continuum index ($\alpha_{\lambda}$), fitted rest-frame flux normalization 
at 1100~\AA,  observed flux  $F_{\lambda}$ at 1300~\AA, and S/N ratios.   We also provide power-law fits to their
\HST/COS spectra (see Section 2.2) and their monochromatic luminosities, $\lambda L_{\lambda}$, at  1100~\AA, 
given by
\begin{equation} 
   \lambda L_{\lambda}  = (1.32\times10^{43}~{\rm erg~s}^{-1}) \left[ \frac {d_L}{100~{\rm Mpc}} \right]^2 
           \left[ \frac {F_{\lambda}} {10^{-14}} \right]  \left[ \frac {\lambda} {1100~{\rm \AA}} \right]  \;.
\end{equation}
We converted flux, $F_{\lambda}$ (in erg~cm$^{-2}$~s$^{-1}$~\AA$^{-1}$) to monochromatic luminosity,
 $L_{\lambda} = 4 \pi d_L^2 F_{\lambda}$, using the luminosity distance, $d_L(z)$, computed for a flat 
 $\Lambda$CDM universe with $H_0 = 70$ km~s$^{-1}$~Mpc$^{-1}$ and density parameters 
 $\Omega_m = 0.275$ and $\Omega_{\Lambda} =0.725$ (Komatsu \etal\ 2011). \\

The redshift distribution of the sample is shown in Figure 1, and the target distribution in accessible rest-frame 
wavelength in Figure 2.  Our sample consists only of those AGN observed with both the G130M (1135--1460~\AA) 
and G160M (1390--1795~\AA) COS gratings, providing the broad wavelength coverage at 17 \kms\ resolution 
needed for our study of the continuum, emission lines, and absorption line blanketing. The COS instrument and 
data acquisition are described by Osterman \etal\ (2011) and Green \etal\ (2012).  We retrieved all COS AGN 
spectra publicly available as of 2013 April 25, but excluded spectra with low signal-to-noise per pixel (S/N $<1$) 
and all BL Lac objects, which are over-represented in the COS archives.  We also excluded three targets with 
abnormal spectra that would complicate the analysis: SDSSJ004222.29-103743.8, which exhibits broad absorption 
lines; SDSSJ135726.27+043541.4, which features a pLLS longward of the COS waveband;  and UGCA 166, which 
Gil de Paz \& Madore (2005) classify as a blue compact dwarf galaxy.  This leaves 159 AGN for analysis.


\begin{figure}[h]
\epsscale{1.2}
\plotone{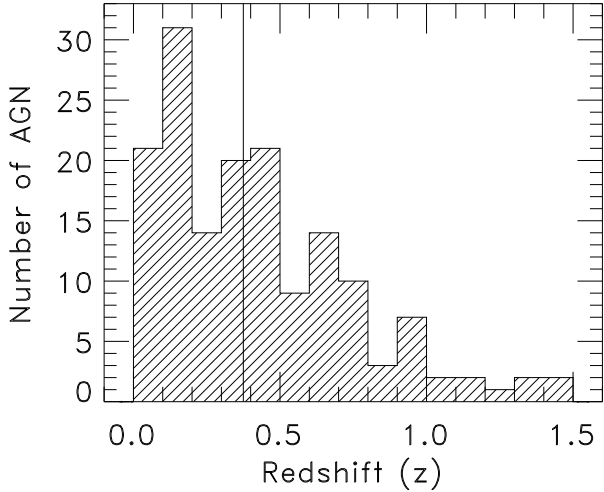}   
\caption{Histogram of the redshifts of the 159 AGN in our COS sample.   Vertical line marks the median 
redshift, $\langle z \rangle \approx 0.37$ of the sample.}
\end{figure}



\begin{figure}[h]
\includegraphics[angle=90,scale=0.32]{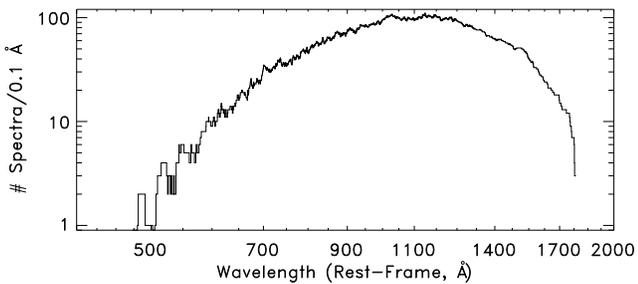}
\caption{Number of AGN targets that contribute to the composite spectrum (see Figure 5) 
as a function of AGN rest-frame wavelength.  Note the rapid decline in targets that probe short 
wavelengths, with ten or fewer AGN probing $\lambda \leq 600$~\AA. }  
\end{figure}


\subsection{Data Acquisition and Processing}

We follow the same procedure as in Paper I for obtaining, reducing, and processing the data.  Below, we briefly 
summarize the procedure and explicitly note improvements or deviations from earlier methods.   Many of these
techniques of coaddition and continuum fitting were also discussed in our IGM survey (Danforth \etal\ 2014).
Of particular interest in this paper are new techniques for identifying LLS and pLLS absorbers, measuring their 
\HI\ column densities, and using that information to correct the continuum.   Our analysis proceeds through the 
following steps:
\begin{enumerate}

\item {\bf Retrieve exposures.}  The CALCOS calibrated exposures were downloaded from the Mikulski Archive for Space 
Telescopes (MAST) and then aligned and co-added using IDL procedures developed by the COS GTO team\footnote{IDL 
routines available at \\ {\tt \url{http://casa.colorado.edu/~danforth/science/cos/costools}} }. Typical wavelength shifts
were a resolution element ($\sim$0.1~\AA) or less, and the co-added flux in each pixel was calculated
as the exposure-weighed mean of the flux in aligned exposures.

\item {\bf Fit spline functions to spectra.}  The raw data contain narrow absorption features that should be excluded 
from the AGN composite spectrum. Identifying and masking each of these features by hand in all of our spectra 
would be extremely tedious. For this reason, we utilize a semi-automated routine that removes narrow absorption 
features and fits the remaining data with a piecewise-continuous function composed of spline functions and 
Legendre polynomial functions. This spline-fitting process involves first splitting the spectra into 5-10~\AA\  segments 
and calculating the average S/N per pixel (flux/error) in each segment. Pixels with S/N less than 1.5 $\sigma$ below 
their segment S/N are rejected from the fitting process to exclude absorption features and regions of increased noise. 
This process is repeated iteratively until there is little change between iterations. The median flux values in the segments 
are then fitted with a spline function. We manually inspect the fits and adjust the identification of rejected regions as 
necessary. Smoothly varying data are well described by this spline-only method. Near broad emission and other 
cusp-like features, short segments of piecewise-continuous Legendre polynomials are preferred. More details on the 
process are given in our IGM survey paper (Danforth \etal\ 2014). 

\item {\bf Deredden spectra.}  We correct the fine-grained data and their corresponding spline fits for Galactic 
reddening, using the empirical mean extinction curve of Fitzpatrick (1999) with a ratio of total-to-selective extinction 
$R_V = A_V / E(B-V) = 3.1$ and color excesses $E(B-V)$ from NED.   In this paper we use values of $E(B-V)$ based 
on dust mapping by Schlegel \etal\ (1998) with a 14$\%$ recalibration by Schlafly \& Finkbeiner  (2011). 
We do not correct for reddening intrinsic to the AGN, although we do not think this could be a substantial effect.
We can probably rule out a large amount of dust (see discussion in Section 3.2).  

 \item {\bf Identify pLLS and LLS absorption.}  In Paper I, we identified pLLS absorption by inspecting the spectra
for flux decrements or Lyman breaks. For this paper we employ a custom computer script that scans each spectrum
for correlated down-pixels at the locations of higher-order Lyman lines of pLLS and LLS absorbers.  First, the script 
divides the spectra by their respective spline fits, normalizing the flux unaffected by IGM absorption to unity. We 
determine the median flux for 15 pixels that have the same relative spacing as the first 15 \HI\ Lyman lines of a 
pLLS with a redshift equal to the source AGN. If there is a pLLS, the median will be much less than unity.  We then 
step one pixel to the left, recalculate the relative spacing of the first 15 Lyman lines at this redshift and the median 
flux for this group of 15 pixels. We repeat this process until we reach the end of the spectrum or a pLLS redshift of zero. 
The script returns a list of redshifts of system candidates to be inspected. When a system is confirmed, 
we measure the equivalent widths of up to the first 12 Lyman lines and fit them to a curve of growth (CoG) to determine 
the column density and Doppler parameter of the system.  In Paper I, we found a total of 17 LLS and pLLS systems in 
8 of the 22 sight lines, and we were sensitive to systems with column density $\log N_{\rm HI} \geq 15.5$. In this paper, 
using our new identification method, we confirm the 17 previously identified systems plus 13 unidentified systems above 
the sensitivity limit $\log N_{\rm HI} \sim15.5$ in the same 22 sight lines from Paper I.  Figure 3 shows examples of pLLS  
identification and continuum restoration. The lowest column density measurement derived from CoG  fitting in this paper is 
$\log N_{\rm HI} \sim13.4$.  We detect  221 systems (7 LLS and 214 pLLS) in 71 of the 159 AGN sight lines, with absorber
redshifts $0.24332 \leq z _a \leq 0.91449$.  These absorbers are listed in Table 2 together with 
our measurements of their redshifts,  \HI\ column densities, and Doppler parameters.  Of the 221 systems, 167 have 
column densities $\log N_{\rm HI} \geq 15.0$ whose distribution in column density is shown in Figure~4.  We only correct 
for these 167 systems in our analysis.  Systems with $\log N_{\rm HI} = 15.5$ and $\log N_{\rm HI} =15.0$ have opacity
that depress the flux immediately blueward of the Lyman limit by less than 2$\%$ and 0.7$\%$ respectively.   Owing to the 
multiple correlated Lyman lines used in this identification technique, our sensitivity is better than the local S/N over most 
of the spectral coverage.  Correcting for the opacity of weaker systems ($\log N_{\rm HI} <15.0$) would have a negligible 
effect on our analysis of AGN continuum, changing the EUV slope of our composite spectrum by only 0.006.

 \item {\bf Restore flux depressed by pLLS and mask unrecoverable flux.} We account for Lyman continuum absorption by 
 measuring the equivalent widths of the first 12 Lyman lines and fitting them with a CoG to estimate the \HI\ column density 
 and Doppler parameter. We use these measurements to correct for the $\nu^{-3}$  opacity shortward of each Lyman edge. 
 We correct only the flux below the Lyman limit.  When a spectrum has pLLS absorption with column density 
 $\log N_{\rm HI} = 15.0-15.9$, we mask the flux between the Lyman limit (911.753~\AA) and Lyman-9 (916.429~\AA) or 
 $\sim$4.7~\AA\ redward in the pLLS rest-frame.  For a pLLS with $\log N_{\rm HI} \geq 15.9$, we mask from the Lyman limit 
 to Lyman-13 (920.963~\AA) or $\sim$9.2~\AA\ redward. When data blueward of the Lyman limit of LLS or pLLS had $S/N < 1$ 
 or did not appear continuous with two or more regions of continuum redward of the Lyman limit, we masked the data.  We also 
 mask regions of the spectra affected by broad absorption from damped \Lya\ systems and  H$_2$ Lyman bands after 
 qualitative visual inspection. The amount of masking varies for individual cases.  Some spectra have one or two gaps of 
 $\leq$10~\AA\ in the data from observations that were not planned with contiguous wavelength coverage over the entire 
 COS-FUV spectral range. These gaps are masked prior to our continuum analysis.

 \item {\bf Shift to rest-frame.}  We shift each spectrum to the rest-frame of the AGN by dividing the wavelength array by 
 $(1+z_{AGN})$. 
 
\item {\bf Mask non-AGN features.}  In every spectrum we exclude Galactic \Lya\ absorption (1215.67~\AA) by 
masking 14~\AA\ on both sides of the line center in the observed frame.  We exclude geocoronal emission lines 
of \NI\ $\lambda1200$ and \OI\ $\lambda1304$ by masking 2~\AA\ on both sides of \NI\ and 5~\AA\ on both sides
of \OI. In five spectra  we masked the absorption due to the \Lya~line of damped \Lya~systems.  

\item {\bf Resample the spectra.}  As in Paper~I and Telfer \etal\ (2002), we resample the spectra 
to uniform 0.1~\AA\ bins. After resampling, each flux pixel corresponds to a new wavelength bin and is equal 
to the mean of the flux in the old pixels that overlap the new bin, weighted by the extent of overlap. 
The error arrays associated with the resampled spectra are determined using a weighting method similar 
to the flux rebinning. See Equations (2) and (3) in Telfer \etal\ (2002) for the rebinning formulae.  

\end{enumerate}


\begin{figure*}[ht]
\includegraphics[angle=90,scale=0.38] {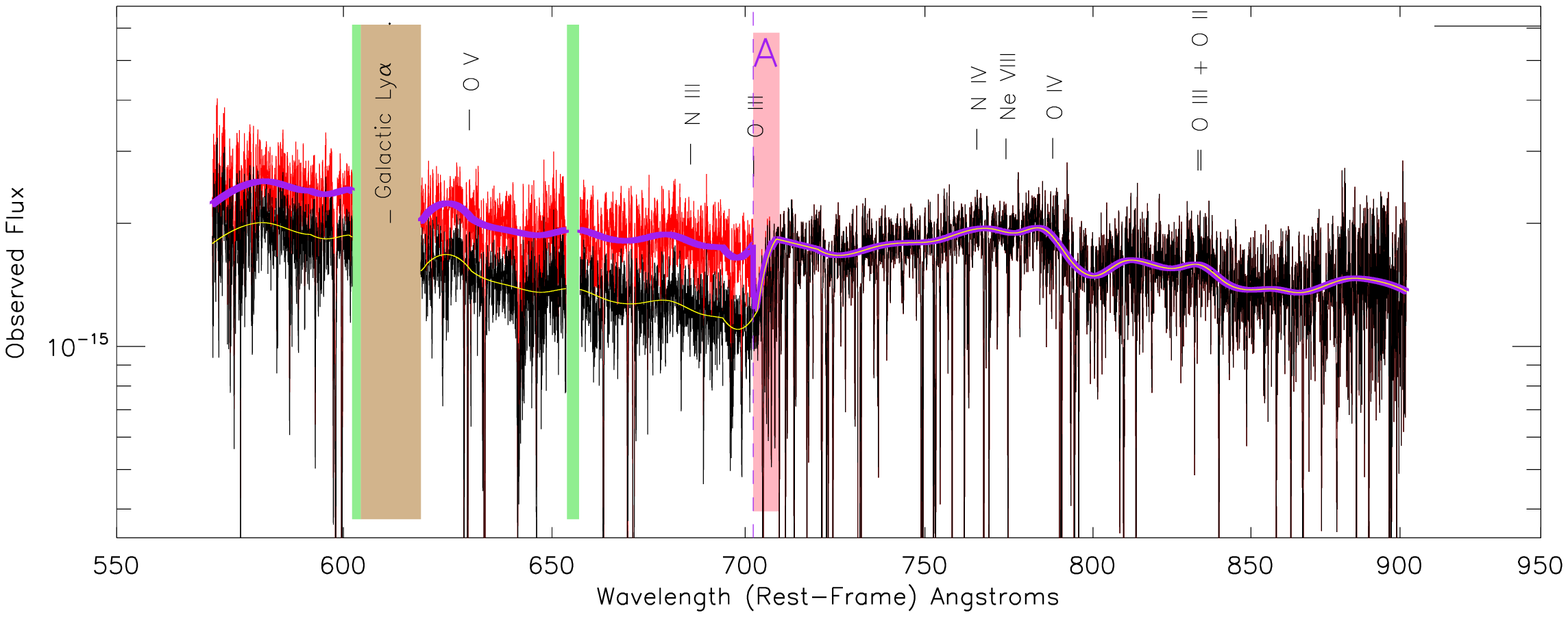}
\includegraphics[angle=90,scale=0.38] {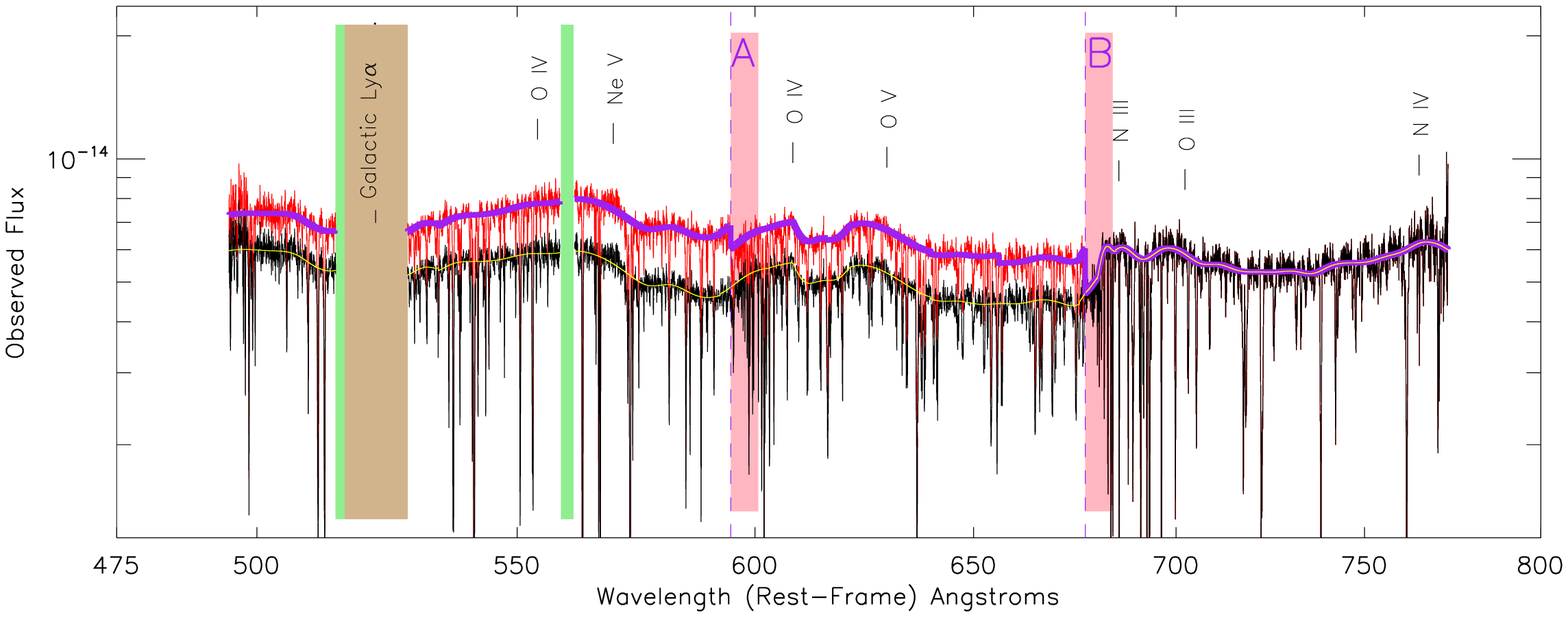}
\includegraphics[angle=90,scale=0.38] {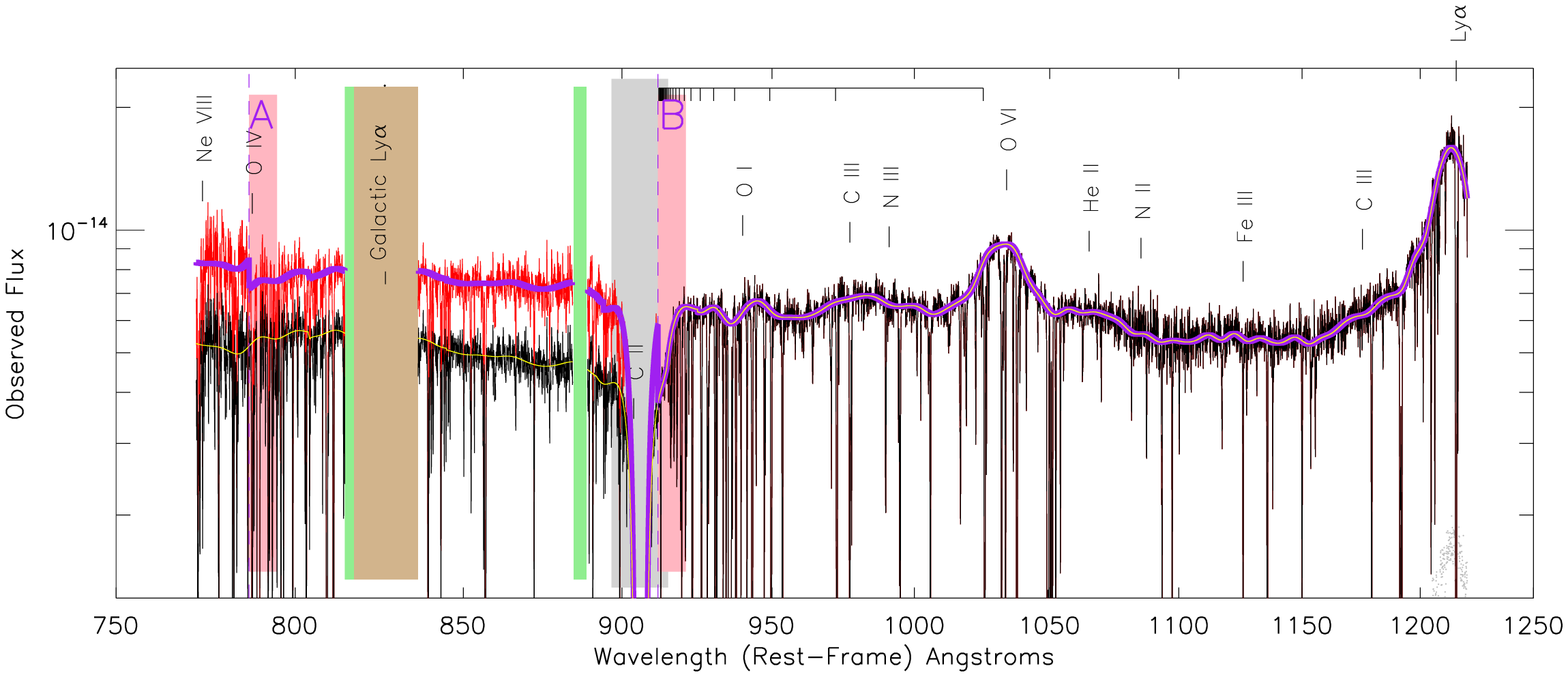}
\includegraphics[angle=90,scale=0.38] {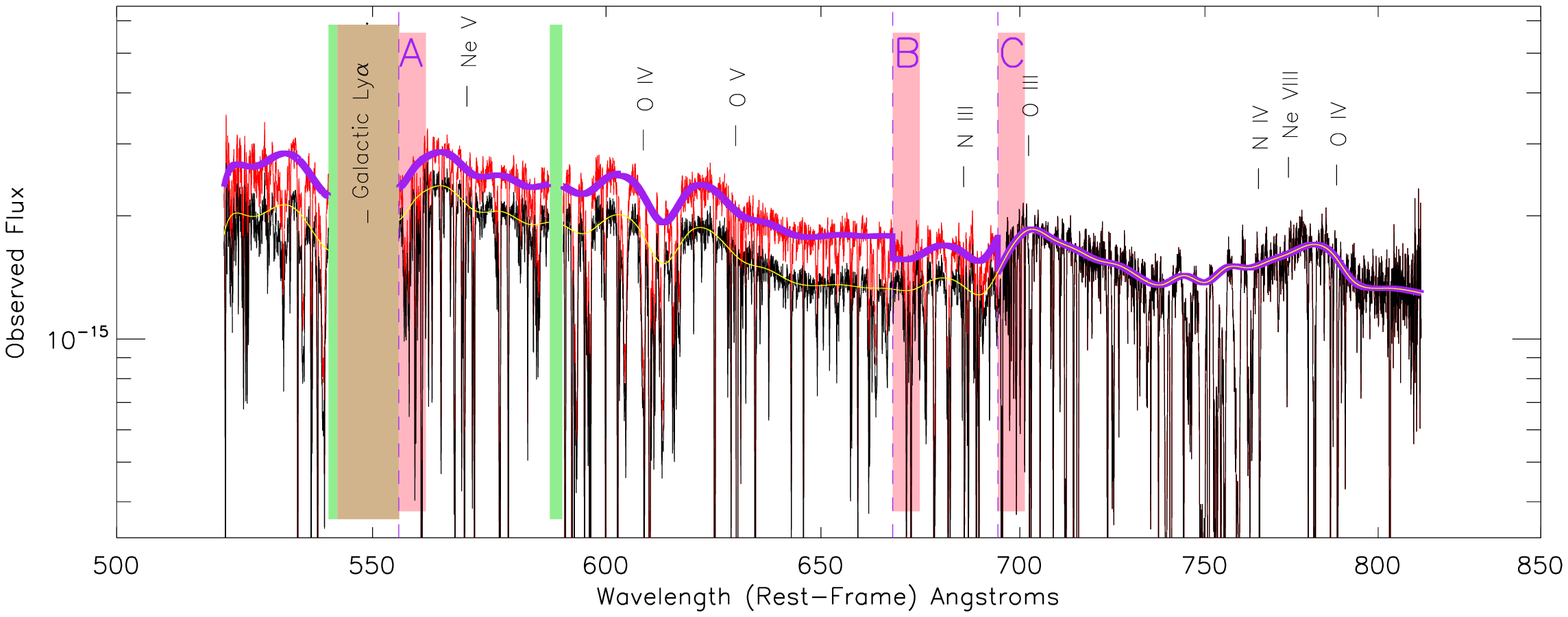}
\caption{Examples of restoring flux depressed by pLLSs:  black line shows flux before 
restoration and red line after restoration; yellow line shows spline fit before restoration and
purple line spline after restoration.  Vertical colored boxes mark data excluded from composite 
and slope measurements:  light brown boxes exclude Galactic \Lya\ absorption; 
light green boxes exclude oxygen airglow emission;  pink boxes exclude absorption from 
LLSs and pLLSs;  gray boxes eliminate large features not intrinsic to AGN emission or 
observational gaps in data.  Panels (top left and right, and bottom left and right) show:  
SDSS J084349.49+411741.6 with absorber A ($z_{\rm LLS} = 0.533$, $\log N_{\rm HI} = 16.77$);
PG 1522+101, with two absorbers:  system A ($z_{\rm LLS} = 0.518$, 
$\log N_{\rm HI} = 16.32$) and system B ($z_{\rm LLS} = 0.729$, $\log N_{\rm HI} = 16.60$);
SDSS J161916.54+334238.4 with system A ($z_{\rm LLS} = 0.269$, $\log N_{\rm HI} = 16.40$) and     
system B ($z_{\rm LLS} = 0.471$, $\log N_{\rm HI} = 16.84$);  
PG 1338+416 with three absorbers:  system A ($z_{\rm LLS} = 0.349$,  $\log N_{\rm HI} = 16.37$), 
system B  ($z_{\rm LLS} = 0.621$,  $\log N_{\rm HI} = 16.30$), and system C ($z_{\rm LLS} = 0.686$,  
$\log N_{\rm HI} = 16.49$).
}
\end{figure*}


\section{OVERALL SAMPLE COMPOSITE SPECTRUM}  
 
\subsection{Composite Construction}

To construct the overall composite spectrum we start by following the bootstrap method of Telfer \etal\ (2002) 
and then adapt the construction technique for our unique dataset. To summarize the bootstrap technique, we 
start near the central region of the output composite spectrum, between 1050~\AA\ and 1150~\AA, and normalize 
the spectra that include the entire central region to have an average flux value of 1 within the central region, 
which we refer to as the ``central continuum window." We then include spectra in sorted order toward shorter 
wavelengths. Lastly, we include the spectra in sorted order toward longer wavelengths. When a spectrum does 
not cover the central continuum window, we normalize it to the partially formed composite by finding the 
weighted-mean normalization constant within multiple emission-line-free continuum windows, calculated using 
Equation (4) of Telfer \etal\ (2002).  We form two independent composite spectra simultaneously: one of the
fine-grained spectra showing the line-blanketing by the \Lya\ forest and interstellar absorption lines, and another 
of the spline fits to the individual spectra.  The spline fits pass over the narrow absorption lines.   

With our unique dataset and spline fits, we adjust the composite construction method in five ways. 
First, with the identification in Paper I of broad emission lines from highly ionized species below 
800~\AA, we were able to restrict the normalization of the spectra at the highest redshifts to two 
narrow regions of continuum-like windows at 660-670~\AA\ and 715-735~\AA.  This is in contrast 
to using all of the flux, including that from emission lines below 800~\AA\ in the calculation of 
the normalization constant, as was done in our initial method. Our second adjustment also limits 
the normalization calculation to regions of continuum, which is our primary interest.  We refine and 
narrow the continuum windows above 800~\AA\ to wavelengths 855-880~\AA, 1090-1105~\AA, 
1140-1155~\AA, 1280-1290~\AA, 1315-1325~\AA, and 1440-1465~\AA.  For our third adjustment 
we choose the region between 855-880~\AA\ as the central continuum window, because it is the largest 
of the narrowed EUV continuum windows with a large number of contributing spectra. Fourth, we note 
that the bootstrapping technique can be sensitive to the ordering in which one includes the spectra, 
especially at the beginning of the process when only a handful of spectra determine the shape of the 
composite.  Therefore, we increase the number of spectra normalized at only the central continuum 
window from 40 to 70 by decreasing the required overlap with the central continuum window from 
100\% to 50\%. Lastly, because we are interested in characterizing the shape of the underlying 
continuum as a power law, we follow the approach of Vanden Berk \etal\ (2001) and combine the 
spectra as a geometric mean, which preserves power-law slopes. We also provide a median-combined 
composite, which preserves the relative shape of the emission features.
 
Below 700 \AA, the number of AGN spectra contributing to each 0.1~\AA\ bin in Figure 2 declines steadily.
Several AGN listed in Table 1 do not appear in this figure because their short-wavelength spectra 
are masked out, owing to LLS absorption, airglow, and pLLS edges.  The final overall sample composite 
spectra (both geometric-mean and median) are presented in two panels of Figure 5, covering rest-frame 
wavelengths from 475-1785~\AA.  In each panel, we show both the fine-grained data with absorption lines
included and the spline-fit continuum composites.  In the geometric-mean spectrum, which we regard as the 
better characterization of the AGN composite, the effects of line-blanketing by the \Lya\ forest can be seen in 
the difference between the spline-fit composite and the real data composite.   Figure 6 shows the optical 
depth, $\tau_{\lambda}$, arising from line-blanketing of the continuum by the \Lya\ forest at 
$\lambda < 1150$~\AA.    We derive optical depths from the difference in fluxes (red and black) in the 
geometric mean composite, shown in the top panel of Figure 5.


\begin{figure}
\epsscale{1.2}
\plotone{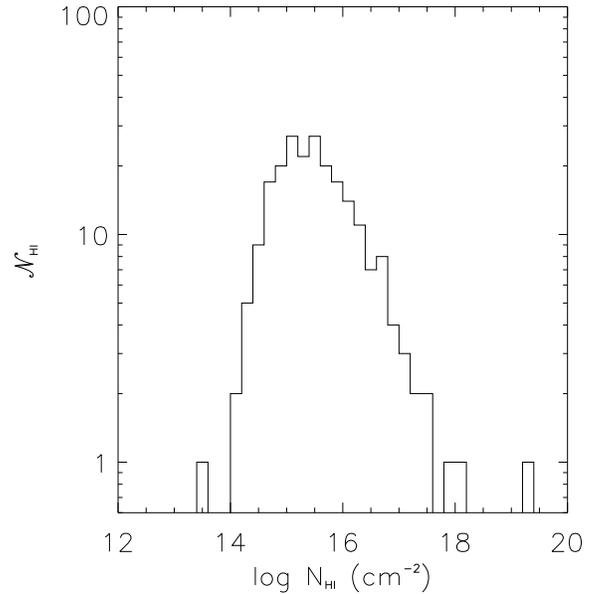}
\caption{Distribution of strong \HI\ absorbers in column density $N_{\rm HI}$, with a range in absorber 
redshifts ($0.24326 < z_a < 0.91449$) accessible to coverage with the COS moderate-resolution 
gratings (G130M and G160M).  Along 71 AGN sight lines at $z_{\rm AGN} > 0.26$, we identified 214 
pLLS with $15.0 \leq \log N_{\rm HI} < 17.2$ and seven LLS with $\log N_{\rm HI} \geq 17.2$. 
 } 
\end{figure}


To characterize the continuum slope of the composite spectrum, we follow the simple approach of 
Vanden Berk \etal\ (2001). We calculate the slope between continuum regions of maximum separation on 
either side of the spectral break, which is clearly present in the composite around 1000~\AA.  Because the 
flux distribution, $F_{\lambda}$, flattens at shorter wavelengths, the two power-law fits pass under the 
observed spectrum and match at the break.  To satisfy this condition, we first calculate the slope of a line 
connecting the minima of the two best continuum regions in log-log space.  We then divide the entire spectrum 
by this line, find the wavelengths of the minima, and calculate the slope of the line that connects the second pair 
of minima.  This results in a line that does not cross the composite. We perform this calculation twice, once in the 
EUV and again in the FUV. We find a mean EUV spectral index $\alpha_{\lambda} = -0.59$ between line-free
windows centered at 724.5~\AA\ and 859~\AA, and mean FUV index $\alpha_{\lambda} = -1.17$ between 
1101~\AA\ and 1449~\AA.   These wavelength indices correspond to frequency indices $\alpha_{\nu}$ of 
$-1.41$ (EUV) and $-0.83$ (FUV).

\subsection{Uncertainties}  

We now discuss sources of random and systematic uncertainty in the composite spectral indices.  
As in Paper I, we fit two power laws to the spline composite spectrum, with indices $\alpha_{\rm FUV}$ 
and $\alpha_{\rm EUV}$, and match them at a break wavelength, which we find to be 
$\lambda_{\rm br} \approx 1000$~\AA, consistent with Paper I and accurate to $\sim50$~\AA.  Although
this gradual break is apparent in the composite, its presence is less clear in the individual spectra, owing 
to the limited spectral range of the COS observations and to strong emission lines of \OVI\ $\lambda1035$ 
and \Lyb\ $\lambda1025$ near the break wavelength.  Because the sample of AGN in Paper I had no targets 
between $0.16 < z < 0.45$, most rest-frame spectra lay either blueward or redward of  the 1000~\AA\ break.  
In our new sample, 55 AGN have redshifts in that range, but we do not distinguish a clear break in individual 
spectra.  With the limited wavelength coverage of COS (G130M, G160M), any single AGN spectrum does not 
have access to the four continuum windows needed to measure two distinct power laws that straddle the break. 
  
To quantify the uncertainty in the fitting of the composite spectrum, we explore the sources of uncertainty described 
by Scott \etal\ (2004), including the effects of intrinsic variations in the shape of the SEDs,  Galactic extinction, 
and formal statistical fitting errors.  As in Paper I, we do not include the effects of intrinsic absorbers or interstellar 
lines, as these absorption lines are easily removed with our moderate-resolution COS spectra.  However, we do 
consider the effects from the strongest systems in the \Lya\ forest. The largest source of uncertainty comes from the 
natural variations in the slope of the contributing spectra. We estimate this uncertainty by selecting 1000 bootstrap
samples with replacement from our sample of 159 AGN spectra.   The resulting distributions of spectral index in 
frequency lead to mean values: $\alpha_{\rm EUV} = -1.41 \pm 0.15$ and $\alpha_{\rm FUV} = -0.83 \pm 0.09$ in 
the EUV and FUV.  Figure 7 shows a montage of spectra for individual AGN, illustrating the wide range of spectral 
slopes and emission-line strengths.     

We also investigated the range of uncertainties arising from UV extinction corrections from two quantities: 
$E(B-V)$ and $R_V$.  We alter the measured $E(B-V)$ by $\pm16$\% ($1\, \sigma$) as reported by Schlegel 
\etal\ (1998).  We deredden the individual spectra with $E(B-V)$ multiplied by 1.16 or 0.84, compile the spectra 
into a composite, and fit the continua.  Over these ranges, we find that  the index $\alpha_{\rm EUV}$ 
changes by ($+0.064, -0.022)$ while $\alpha_{\rm FUV}$  changes by $(+0.046, -0.023)$.    Next, we estimate 
the sensitivity to deviations from the canonical value $R_V = 3.1$, which Clayton, Cardelli, \& Mathis (1988) 
found to vary from $R_V = 2.5$ to $R_V = 5.5$.  We follow Scott \etal\ (2004) and deredden individual spectra 
with $R_V = 2.8$ and $R_V = 4.0$ and compiling the spectra into composites. We find that $\alpha_{\rm EUV}$ 
changes by $(+0.041, -0.051)$ and $\alpha_{\rm FUV}$ by $(+0.032, -0.059)$.  We estimate the uncertainties 
arising from correcting pLLS absorption of strength $\log N_{\rm HI} \geq 15.0$ by altering the measured column 
densities by $\pm 1\, \sigma$ as reported in Table 2.  We find that $\alpha_{\rm EUV}$ changes by
$(+0.037, -0.010)$ and $\alpha_{\rm FUV}$ by $(+0.011, -0.011)$. The formal statistical errors for the 
spectral indices are negligible ($<0.001$),  owing to the high S/N ratio of our composite spectra.  Thus, 
we do not include them in the final quoted uncertainties.  We add the random uncertainties of cosmic variance 
with the systematic effects of correcting for extinction in quadrature and estimate the total uncertainties 
to be $\pm 0.15$ for $\alpha_{\rm EUV}$ and $\pm 0.09$ for $\alpha_{\rm FUV}$.  

As noted earlier, we do not correct for AGN dust, but small amounts could be present, as long as they do not
produce a strong turnover in the far-UV fluxes.  We can rule out a large amount of intrinsic extinction, if it obeys 
a selective extinction law, $A(\lambda)/A_V$, that rises steeply at short (UV) wavelengths.  Otherwise, we would 
see steep curvature in the rest-frame EUV, rather than a power law.

\subsection{Comparison to Other Composite Spectra} 

Ultraviolet spectra of AGN have been surveyed by many previous telescopes, including the {\it International 
Ultraviolet Explorer} (O'Brien \etal\ 1988) and the \HST\ Faint Object Spectrograph (Zheng \etal\ 1997).  More recent 
AGN composite spectra were constructed from data taken with \HST/FOS+STIS (Telfer \etal\ 2002),  \FUSE\ (Scott \etal\ 
2004), and \HST/COS (Shull \etal\ 2012 and this paper).   Figure 8 compares the \HST/COS composites with previous 
studies with \HST/FOS+STIS and \FUSE.   Our current COS survey finds essentially  the same EUV spectral index, 
$\alpha_{\nu} = -1.41 \pm 0.15$, as found in Paper~I, $\alpha_{\nu} = -1.41 \pm 0.22$, but with better statistics and 
coverage to shorter wavelengths (below 500~\AA).   This consistency is reassuring, as our current composite includes 
159 AGN spectra, compared with 22 AGN in the initial COS study (Paper~I).  The \HST/COS EUV index, 
$\alpha_{\nu} \approx  -1.4$ is slightly harder than the \HST/FOS+STIS value, $\alpha_{\nu}  = -1.57\pm0.17$, found 
by Telfer \etal\  (2002) for 39 radio-quiet AGN.  However, both \HST\ surveys found indices steeper than the \FUSE\ 
slope, $\alpha_{\nu} = -0.56^{+0.38}_{-0.28}$ (Scott \etal\ 2004), a puzzling discrepancy that we now investigate. 

The differences between continuum slopes found in \FUSE\ and COS composite spectra are likely to arise from 
four general factors: 
(1) continuum placement beneath prominent EUV emission lines; 
(2) line blanketing by the \Lya\ forest and stronger (pLLS) absorbers;  
(3) continuum windows that span an intrinsically curved AGN spectrum;  and
(4) possible correlation of slope and AGN luminosity.  
High-S/N spectra at the moderate resolution of COS (G130M/G160M) are critical for identifying the underlying 
continuum near the strong EUV emission lines of \OIII, \OIV, and \OV\ and the  \NeVIII\ doublet ($\lambda\lambda 
770,780$).   The COS spectral resolution also allows us to fit over the narrow \HI\ absorbers in the \Lya\ forest (factor~2) 
and restore the continuum absorbed by the stronger systems (LLS and pLLS).  Factor 3 is a more subtle effect, but 
it may be the most important.   The COS wavelength coverage (1135--1795~\AA) is broader than that of \FUSE\ 
(912--1189~\AA), and it provides line-free continuum windows above and below 1100~\AA, spanning an intrinsically 
curved SED.   This allows us to construct a two-component spectrum with indices
 $\alpha_{\nu} = -1.41 \pm 0.15$ in the EUV (500-1000~\AA) and $\alpha_{\nu} = -0.83 \pm 0.09$ in the FUV (1200-2000~\AA) 
 with a break at $\lambda_{\rm br} \approx 1000\pm25$~\AA.    Many of the COS sight lines observe higher-redshift AGN
that sample different regions of the SED than those of \FUSE.  
Shortward of 912~\AA, we place the continuum below a number of prominent emission lines, using nearly line-free 
continuum windows at $665\pm5$~\AA,  $725\pm10$~\AA, and $870\pm10$~\AA.   Factor~4 refers to possible selection 
effects of AGN luminosity with redshift.   Previous samples used targets at a variety of redshifts and luminosities, 
observed with different spectral resolution, FUV throughput, and instruments.  All of the UV composite spectra (\HST\ and
\FUSE) are based on the available UV-bright targets (Type-1 Seyferts and QSOs) studied with \IUE\ and the {\it Galaxy 
Evolution Explorer} (\Galex).  Most of these AGN were chosen as background sources for studies of IGM, CGM, and 
Galactic halo gas.  Although these targets are not a complete, flux-limited sample of the AGN luminosity function 
(e.g., Barger \& Cowie 2010),  they probably are representative of UV-bright QSOs, at least at redshifts $z < 0.4$.  

Figure 9 compares the average AGN redshift per wavelength bin for the COS and \FUSE\ surveys, overlaid on the line-free 
continuum windows.   Evidently, the COS targets are at systematically higher redshift, and their wavelength coverage is
broader than that of \FUSE.  The average AGN luminosity also differs, longward and shortward of the break.  At 
$\lambda \approx 800$~\AA, the COS and \FUSE\ composites are both probing similar luminosities.  As shown in the top 
panel of Figure 9, the two spectral slopes are fairly similar between 650--1000~\AA, and the only difference comes from the 
sudden decline in \FUSE\  fluxes between 1090--1140~\AA.  Lacking the longer-wavelength continuum windows, the \FUSE\ 
spectra were unable to fit the break in spectral slope at longer wavelengths.  
Figure 10 shows the distributions of spectral index $\alpha_{\lambda}$ and the effects of the available continuum windows 
falling longward or shortward of the 1000~\AA\ break.  The two-power-law fits possible with COS data allow us to measure
the spectral curvature and distinguish between FUV and EUV slopes.  This was not done with the \FUSE\ composite fits.  

In summary, we believe the \HST/COS composite spectra are superior owing to their higher spectral resolution (G130M and 
G160M gratings) allowing us to resolve and mask out the \Lya\ forest and restore the continuum from stronger (LLS and pLLS)
absorbers.   The higher S/N of the COS spectra allow us to identify and resolve prominent UV/EUV emission lines and fit a 
more accurate underlying continuum.  As shown in Figures 2 and 7, the COS composite still contains fewer than 10 AGN at 
$z > 1$ that probe the rest-frame continuum at  $\lambda < 600$~\AA.   These numbers are larger than in the earlier surveys, 
but the small  sample means that the composite spectrum remains uncertain at the shortest  wavelengths.

\subsection{Trends with Redshift, AGN Type, and Luminosity}

As in Paper I, we explore trends within the \HST/COS AGN sample by constructing composite spectra based on various 
parameters and subsamples.  Figure 11 shows the distributions of index $\alpha_{\lambda}$ in redshift, AGN activity type,  
Galactic foreground reddening, and monochromatic (1100~\AA) luminosity.   In each panel, two horizontal lines denote the 
sample-mean values:
$\langle \alpha_{\lambda} \rangle = -0.59$ for the rest-frame EUV (500-1000~\AA) band and 
$\langle \alpha_{\lambda} \rangle = -1.17$  for the rest-frame FUV (1200-2000~\AA) band.
The spectral indices extend over a wide range of AGN luminosities and activity types, with no 
obvious trend or correlation.   Galactic reddening does not appear to produce any difference in the index.  There may be 
subtle trends in the distribution of $\alpha_{\lambda}$ with redshift, because we are observing the rest-frame flux from an 
intrinsically curved  spectral energy distribution (SED).   
At low redshift ($z<0.25$) there are many AGN with steep slopes, $\alpha_{\lambda}  <-1.5$, indicating hard UV spectra.  
However, only seven AGN have spectra with $\alpha_{\lambda}  <-1.5$.  At higher redshift ($z>0.5$) there are few AGN 
with slopes  $\alpha_{\lambda} <-1.5$, and the survey contains few AGN at the most extreme redshifts ($z > 1$).   Many 
more have soft spectra with slopes $\alpha_{\lambda} >-0.5$.

 \subsection{Softened UV Spectra and Continuum Edges} 
  
Accretion disk (AD) model spectra have recently been investigated by a number of groups (Davis \etal\ 2007;  Laor \& Davis  
2011; Done \etal\ 2012; Slone \& Netzer 2012) with a goal of comparing to UV and EUV spectra.   The observed far-UV spectral 
turnover at $\lambda < 1000$~\AA\  limits the maximal disk temperature to $T_{\rm max} \approx 50,000$~K.  Model 
atmospheres computed with the {\it TLUSTY} code (Hubeny \etal\ 2001) and including winds driven from inner regions of 
the disk predict a spectral break near 1000~\AA, arising from the Lyman edge (912~\AA) and wind-truncation of the hot 
inner part of the disk (Laor \& Davis 2014).  In a standard multi-temperature accretion-disk models with blackbody spectra 
in annular rings (Pringle 1981) the radial temperature distribution scales as $T(r) \propto (M_{\rm BH} \dot{M} / r^3)^{1/4}$.  
In their models of wind-ejecting disks,  Slone \& Netzer (2012) suggest that the spectral shape is governed by the radial profile 
of $\dot{M}(r)$, and the radius $r_{1/2}$ where half the disk mass has been ejected.  Their observational predictions 
are based on the removal of hot accreting gas from the inner regions of the AD and accompanying removal of energy from the 
UV-emitting portions of the SED.   A large mass accretion rate throughout the AD produces higher luminosities and shifts the 
SED to shorter (UV) wavelengths. The closer $r_{1/2}$ comes to the innermost stable circular orbit, $r_{\rm ISCO}$, the more 
FUV and EUV radiation will be emitted.  

Slone \& Netzer (2012) used the sensitivity of the EUV spectral index, $\alpha_{456-912}$ between 456--912~\AA, to constrain 
accretion properties outside $R_{\rm ISCO}$, the radius of the innermost stable circular orbit around a black hole (see their 
Figure 7).  Their model is governed by mass accretion rates, ${\dot M}_{\rm in}$ and  ${\dot M}_{\rm out}$, at the inner and outer 
disk radii of the disk, relative to the Eddington accretion  rate $\dot{M}_{\rm edd}$ and luminosity $L/L_{\rm edd}$ relative to the
Eddington luminosity, $L_{\rm edd}$.  From the observed EUV spectral index, $\alpha_{456-912} \approx -1.4$, we constrain the 
mean AGN accretion rate and luminosity to values $\dot{M}_{\rm in} / \dot{M}_{\rm edd} < 0.1$ and $L/L_{\rm edd} < 0.2$.   
We caution that these inferences are subject to the validity of accretion-disk model atmospheres, including effects of external
irradiation, uncertainty in where energy is being deposited, and the role of magnetic field energy dissipation.  In addition,
disk photospheres may differ from those of hot stars, with spatially variable $\tau = 1$ surfaces.   
  
 Laor \& Davis (2014) explore similar disk-truncation models, solving for the radial structure of a disk with mass loss.
 They find that the wind mass loss rate, $\dot{M}_{\rm wind}$, becomes comparable to the total accretion rate $\dot{M}$
 at radii a few tens of gravitational radii, $(GM/c^2)$.  Line-driven winds set a cap of $T_{\rm max} < 10^5$~K on their 
 disks, which in most cases are truncated well outside the ISCO radius.  These models are consistent with the observed
 SED turnover at $\lambda < 1000$~\AA\ that is weakly dependent on luminosity $L$ and black-hole mass $M_{\rm BH}$.  
 Their models of line-driven winds also cap AD effective temperatures, $T_{\rm eff} < 10^5$~K.   The UV spectral turnover
 is produced by both an \HI\  Lyman edge and the limit on disk temperature.  
 
 Standard models of accretion disk atmospheres are predicted to exhibit  \HI\ and \HeI\ continuum edges at 912~\AA\
 and 504~\AA, respectively.   This issue and the EUV (soft X-ray) spectra of accretion disks have been discussed  by 
 many authors (e.g., Kolykhalov \& Sunyaev 1984; Koratkar \& Blaes 1999; Done \etal\ 2012).   The absence of any
 continuum absorption at 912~\AA\ in the composite spectrum was noted in Paper~I, where we set an optical depth limit 
 of $\tau_{\rm HI} < 0.03$.  From the 159-AGN composite (see Figures 5 and 8) our limit is now $\tau_{\rm HI} < 0.01$ 
 derived from the flux around 914.5~\AA\ and 910.5~\AA.    The limit for the \HeI\ edge at 504~\AA\ is less certain because 
 of the difficulty in fitting the local continuum under neighboring broad EUV emission lines.  However, from the general
 continuum shape between 480-520~\AA, we can limit the \HeI\ continuum optical depth to $\tau_{\rm HeI} < 0.1$. 
 Additional COS/G140L data now being acquired toward 11 AGN at redshifts $1.5 \leq z \leq 2.2$ probe the rest-frame c
 ontinua at $\lambda < 400$~\AA\ with good spectral coverage at the 504~\AA\ edge.  We continue to see no \HeI\ 
 continuum edge.

\section{DISCUSSION AND CONCLUSIONS}  

We now summarize the results and implications of our \HST/COS survey of AGN spectral distributions in the AGN rest-frame 
FUV and EUV. Using spectra of 159 AGN taken with \HST/COS G130M and G160M gratings, we constructed a 2-component
composite spectrum n the EUV (500-1000~\AA) and  FUV (1200-2000~\AA).  These two spectral fits match at a break wavelength
$\lambda_{\rm br} \approx 1000$~\AA, below which the SED steepens to $F_{\nu} \propto \nu^{-1.41}$.  The EUV index
is the same as found in Paper I, but with smaller error bars.   It is slightly harder than the index, $\alpha_{\nu} = -1.57 \pm 0.17$, 
found from the HST/FOS+STIS survey (Telfer \etal\ 2002) for radio-quiet AGN, but much softer than the index, 
$\alpha_{\nu} = -0.56^{+0.38}_{-0.28}$, from the \FUSE\ survey (Scott \etal\ 2004).  These composite spectra are based
on small numbers of AGN with redshifts ($z \geq 1$) sufficient to probe below 600~\AA.   However, the \HST/COS survey 
provides a superior measure of the true underlying continuum.   Our G130M/G160M data have sufficient spectral resolution 
and signal to noise to mask out narrow lines from the \Lya\ forest and restore the continuum from stronger (LLS and pLLS) 
absorbers.  We also fit the continuum below the prominent broad EUV emission lines using nearly line-free continuum windows 
at $665\pm5$~\AA,  $725\pm10$~\AA, and $870\pm10$~\AA.  

\vspace{.3cm}

\noindent
Our primary conclusions are as follows:  
\begin{enumerate}

\item The HST/COS composite spectrum follows a flux distribution with $F_{\nu} \propto \nu^{-0.83 \pm0.09}$ 
    for AGN rest-frame wavelengths 1200-2000~\AA\ and $F_{\nu} \propto \nu^{-1.41\pm0.15}$ for 500-1000~\AA.   
    This  EUV spectral index is slightly harder than that used in recent simulations (Haardt \& Madau 2012) of IGM 
    photoionization and photoelectric heating.
    
\item Individual spectra of the 159 AGN surveyed exhibit a wide range of spectral indices in the EUV,  with typical
    values between $-2 \leq \alpha_{\nu} \leq 0$.  These indices are {\it local} slopes and not characteristic of the 
    spectral energy distribution over the full UV/EUV band,  
   
\item The composite SED exhibits a turnover at $\lambda < 1000$~\AA, characteristic of accretion disk models
     in which the maximum temperature $T_{\rm max} < 10^5$~K and the inner disk is truncated by line-driven winds. 
   
\item We see no continuum edges of \HI\ (912~\AA) or \HeI\ (504~\AA), with optical depth limits $\tau_{\rm HI} < 0.01$
    and $\tau_{\rm HeI} < 0.1$.   The absence of these edges suggests that accretion disk atmospheres differ
    from those of hot stars, because of external irradiation or inverted temperature structures arising
    from magnetic energy dissipation.

\item We find no obvious correlations of the EUV spectral index with interstellar reddening, AGN type, redshift,
  or luminosity ($\lambda L_{\lambda}$ at 1100~\AA).   Such trends are difficult to pick out, because the observable
  \HST/COS (G130M/G160M) wavelength band (1135-1795~\AA) covers different portions of the SED over the AGN 
  redshifts $0.001 < z < 1.476$) in our sample.  The quoted indices, $\alpha_{\lambda}$ are {\it local} slopes that
  fall either in the FUV or EUV depending on AGN redshift.  
   
\item  The mean EUV slopes, compared to models of wind-truncated thin accretion disks, constrain the mean 
   accretion rate in the inner disk and the AGN luminosity to values $\dot{M}_{\rm in} / \dot{M}_{\rm edd} < 0.1$ and 
   $L/L_{\rm edd} < 0.2$ relative to their Eddington rates. 
   
\end{enumerate}

\vspace{0.5cm}

The order-of-magnitude improvement in sensitivity offered by COS over previous spectrographs has greatly
increased the number of targets available for moderate-resolution UV spectroscopy.   Some of these spectra have 
S/N below the threshold chosen for this survey, and many are low-resolution (G140L) rather than G130M/G160M 
used here.  Nevertheless, some of these archival spectra will provide EUV coverage down to 500~\AA\ (with 
$\sim$40 AGN) and to 912~\AA\ (with $\sim$100 AGN).    Additional data at AGN rest wavelengths 400-500~\AA\ 
would be helpful in fitting the SED deeper into the EUV, where fewer than ten AGN sight lines have been probed 
to date with COS/G130M.   Currently, our composite spectrum includes 10 AGN that contribute at 
$\lambda \leq 600$~\AA\  but only two AGN at  $\lambda < 500$~\AA.  As noted in Paper~I,  one can explore 
even shorter rest-frame wavelengths (304--500~\AA) using the sample of ``\HeII\ quasars" (Worseck \etal\ 2011; 
Syphers \etal\ 2011; Shull \etal\ 2010) that probe the \HeII\  epoch of reionization at $z \approx 2.5-3.5$.  
 In Hubble Cycle 21, we are observing 11 new AGN targets at $z = 1.45$ to $z = 2.13$, using the lower-resolution 
 (G140L) grating.  The first ten of these spectra have now been acquired.  After reduction, they should improve the 
 accuracy of the composite spectrum down to 400~\AA\ and provide more sight lines that cover the \HeI\ 504~\AA\ 
 continuum edge.  Our intent is to create a composite AGN spectrum between 350~\AA\ and 1800~\AA, using 
 \HST/COS archival spectra of AGN with a variety of types and  luminosities.


\acknowledgments

\noindent
We thank the COS/GTO team for help on the calibration and verification of  COS data.  We acknowledge helpful
discussions with Shane Davis, Ari Laor, and Jim Pringle on accretion disk models and thank the referee for
helpful comments that encouraged us to explore the differences between COS and \FUSE\ composite spectra.
This research was supported by NASA grants NNX08-AC14G and NAS5-98043 and the Astrophysical Theory Program
(NNX07-AG77G from NASA) at the University of Colorado Boulder.  JMS thanks the Institute of Astronomy, Cambridge 
University, for their stimulating scientific atmosphere and support through the Sackler Visitor Program.


\clearpage


\begin{figure}[h]
\includegraphics[angle=90,scale=0.7]{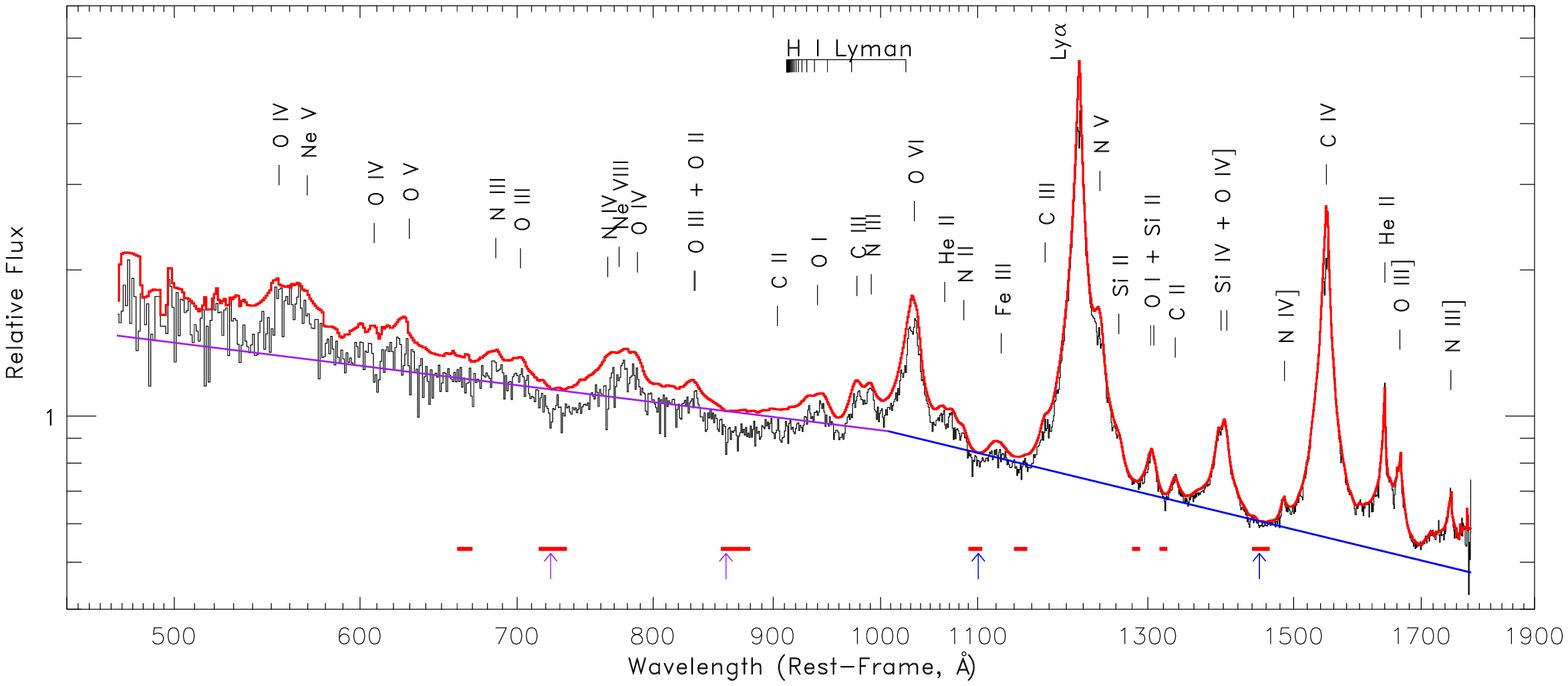}
\includegraphics[angle=90,scale=0.7]{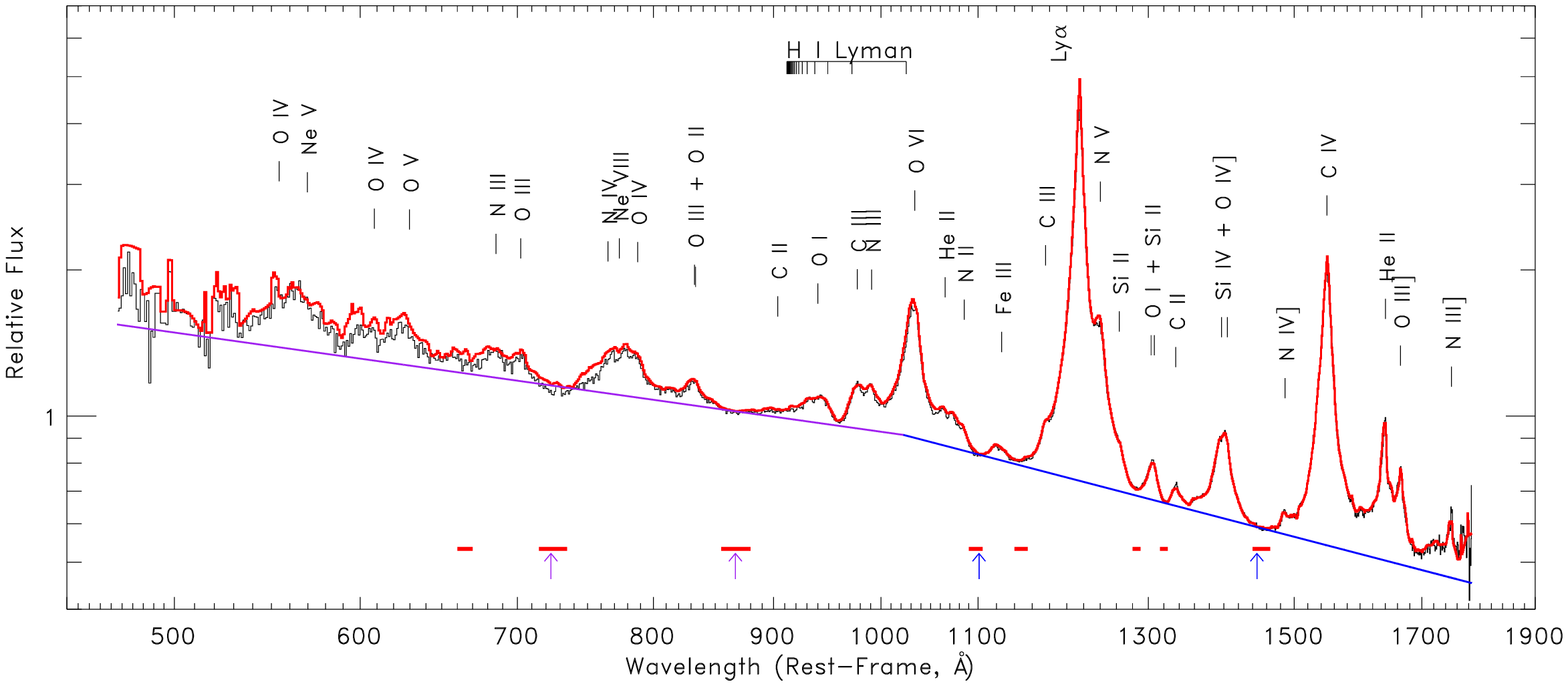}
\caption{\small Composite spectra with rest-frame wavelengths 465--1750~\AA\ made from 159 AGN 
with redshifts $0.001 \leq z_{\rm AGN} \leq 1.476$,  resampled to 0.1~\AA, plotted in 1~\AA\ bins, 
normalized to unit flux at 1100~\AA, and showing broad FUV and EUV emission lines atop a power-law continuum.   
Eight continuum windows are shown as small red boxes along bottom.  Composite data are shown in black; 
red curve is composite of individual spline fits. 
(Top) Geometric-mean-combined  \HST/COS spectrum with frequency distribution, 
$F_{\nu} \propto \nu^{\alpha_{\nu}}$, with break at  $\lambda_{\rm br} \approx 1000$~\AA\
and  spectral indices $\alpha_{\nu} = -1.41 \pm 0.15$ (EUV, $\lambda < 1000$~\AA) and
 $\alpha_{\nu} = -0.83 \pm 0.09$ (FUV, $\lambda  >1200$~\AA).  
(Bottom)  Median-combined composite, with break at  $\lambda_{\rm br} \approx 1025$~\AA\
and spectral indices $\alpha_{\nu} = -1.32 \pm 0.15$ (EUV) and $\alpha_{\nu} = -0.74 \pm 0.09$ (FUV).  
 }  
\end{figure}



\begin{figure}
\epsscale{0.7}
\plotone{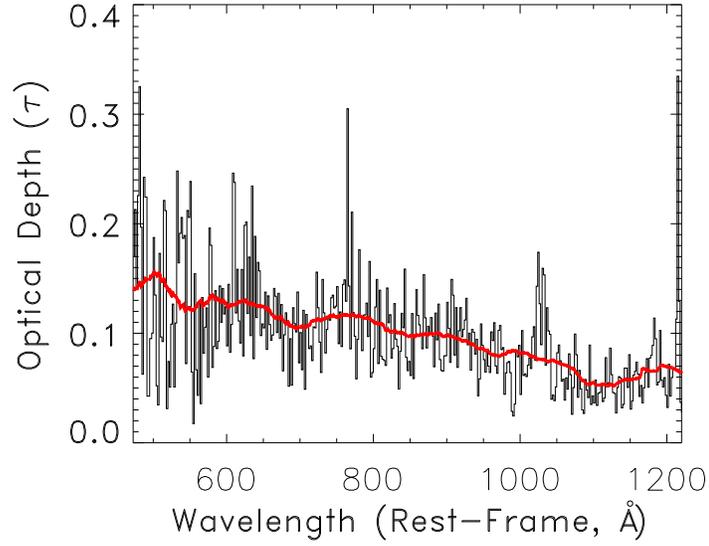} 
\caption{Optical depth arising from continuum line-blanketing by the \Lya\ forest, calculated from the 
difference in fluxes (top panel of Figure 5) between spline-fit continuum and line-blanketed data, and
binned by 20 pixels of 0.1~\AA\ width.    Red overplot shows optical depth, smoothed over 1000 pixels. } 

\end{figure}



\begin{figure}[h]
\includegraphics[angle=90,scale=0.7]{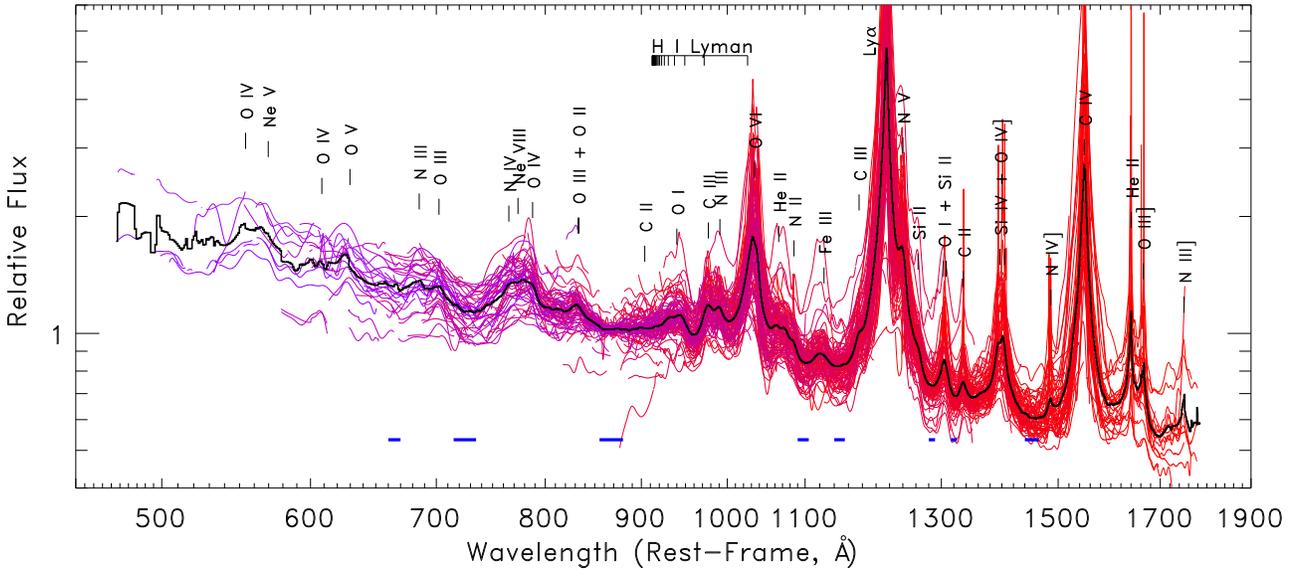} 
\caption{Montage of spectra of the 159 individual AGN that go into the composite spectrum (solid line), 
showing the range of spectral slopes and variations in emission-line strengths.  The line-free
continuum windows are shown as blue bars along bottom. }
\end{figure}



\begin{figure}[h]
\includegraphics[angle=90,scale=0.69]{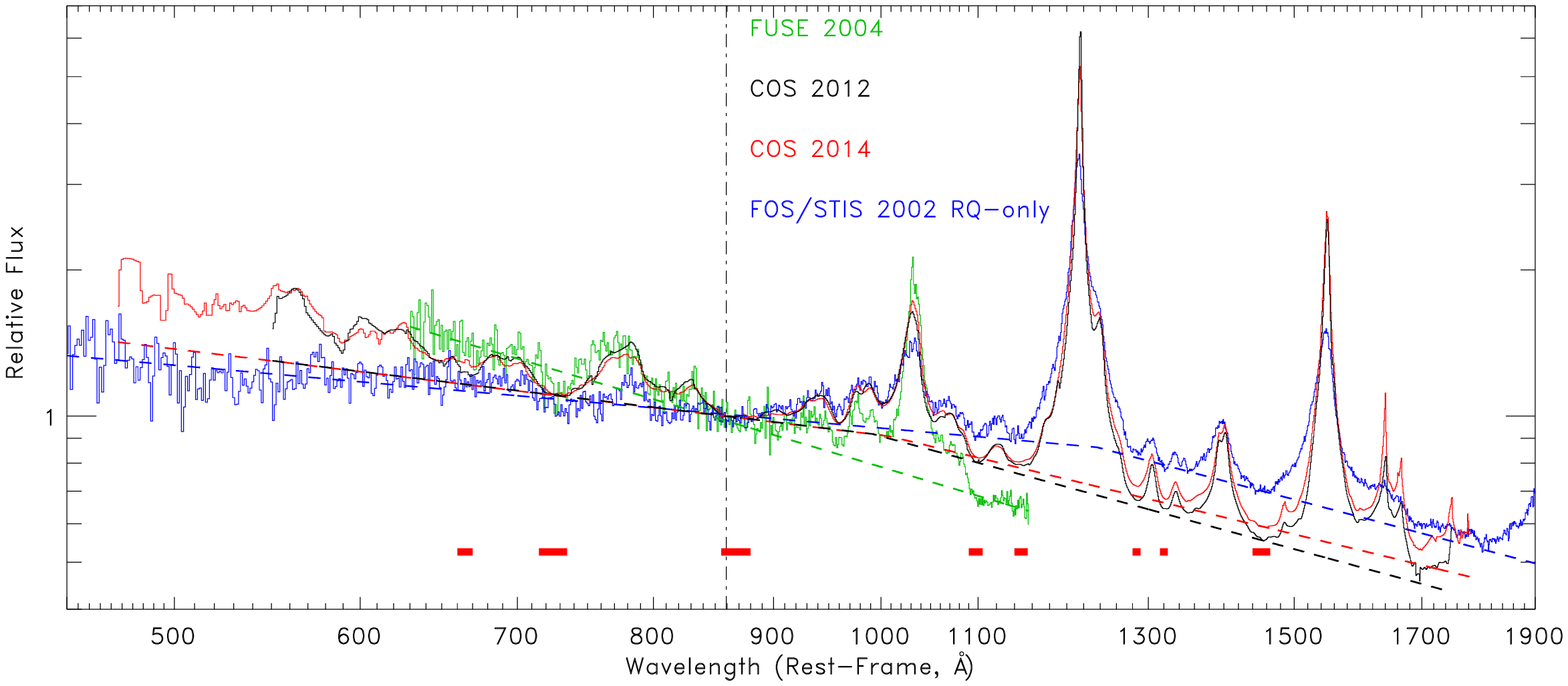}
\includegraphics[angle=90,scale=0.69]{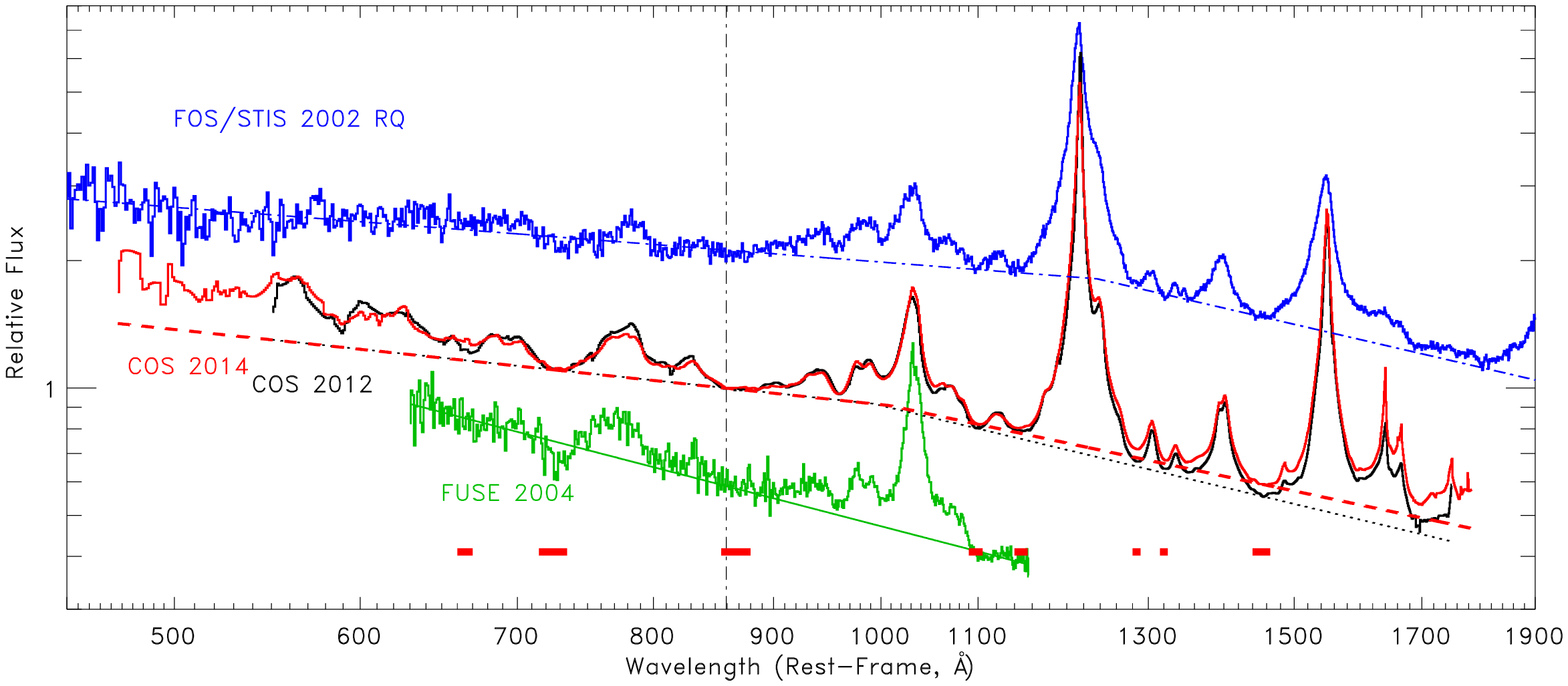}
\caption{\small Comparison of the AGN composite spectra from FUSE (Scott \etal\ 2004), HST/FOS+STIS (Telfer \etal\ 
2004), and two surveys with HST/COS (Shull \etal\ 2012; Stevans \etal\ 2014).  {\it (Top.)} Composite spectra, $F_{\lambda}$, 
all normalized at the 860~\AA\ line-free continuum window.  Dashed lines show underlying power-law continua.  The COS (2012, 2014) 
composite spectra, $F_{\nu} \propto \nu^{\alpha_{\nu}}$, have essentially the same EUV spectral index, 
$\alpha_{\nu} = -1.41\pm0.15$.  FOS/STIS spectrum has $\alpha_{\nu} = -1.57 \pm 0.17$ 
for 39 radio-quiet QSOs.   FUSE spectrum is harder with $\alpha_{\nu} = -0.56^{+0.38}_{-0.28}$ based on lower-redshift  AGN.  
{\it (Bottom.)} Spectra offset vertically for clarity.  The different slopes arise from fitting the continuum beneath prominent broad 
EUV emission lines (\NeVIII, \OIII, \OIV, \OV) and HST/COS  wavelength coverage spanning an intrinsically curved AGN spectrum.   
Access to line-free continuum windows above and below 1100~\AA\ (red bars along bottom) allows us to fit different FUV and 
EUV continuum slopes with a break at $\lambda_{\rm br} \approx 1000\pm25$~\AA.
}
\end{figure}



\begin{figure}[h]
\includegraphics[angle=0,scale=1.0]{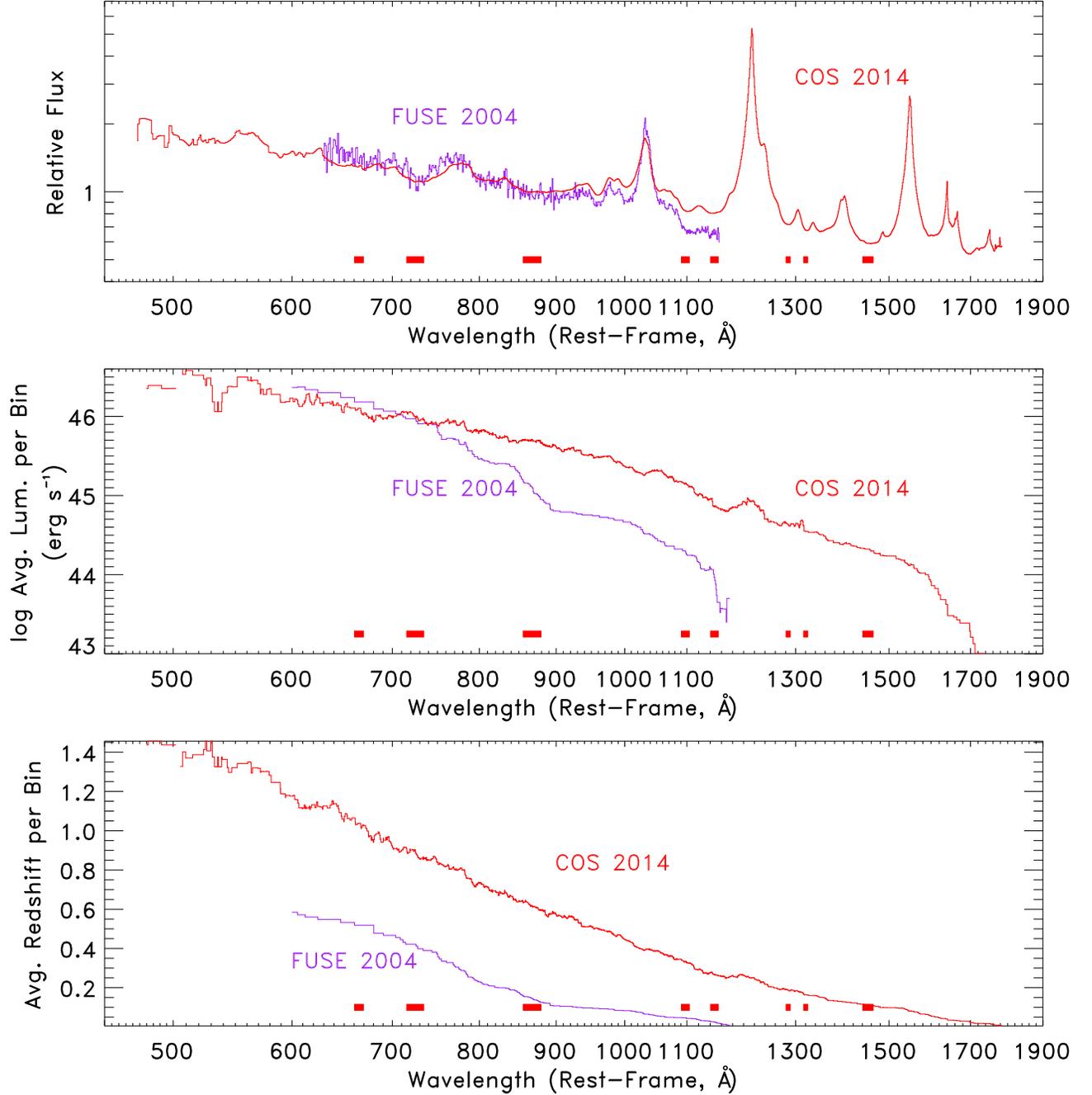}
\caption{\small  Comparison of composite spectra from COS and \FUSE, with line-free continuum windows shown as 
red boxes at 660-670~\AA, 720-730~\AA,  855-880~\AA, 1090-1105~\AA, 1140-1155~\AA, 1280-1290~\AA, 1315-1325~\AA, 
and 1440-1465~\AA.  
  {\it (Top).}   Aligned composite spectra of COS and \FUSE,  normalized to 1 at  860~\AA.  Spectra agree at short wavelengths 
  ($\lambda \leq 1000$~\AA) but the \FUSE\ fluxes show a sudden drop-off longward of 1100~\AA\ in continuum windows at 
  1100~\AA\ and 1145~\AA.  
  {\it (Middle).}  Average AGN luminosity per wavelength bin.   We plot the geometric mean luminosity, $\lambda L_{\lambda}$,
     at 1000~\AA.   
  {\it (Bottom).}  Average redshift of AGN per wavelength bin.
The  \FUSE\ survey samples AGN at lower redshift than COS, with AGN luminosities comparable at $\lambda < 800$~\AA, but 
much lower at $\lambda > 900$~\AA.  
 }
 \end{figure}



\begin{figure}
\epsscale{1.1} 
\plotone{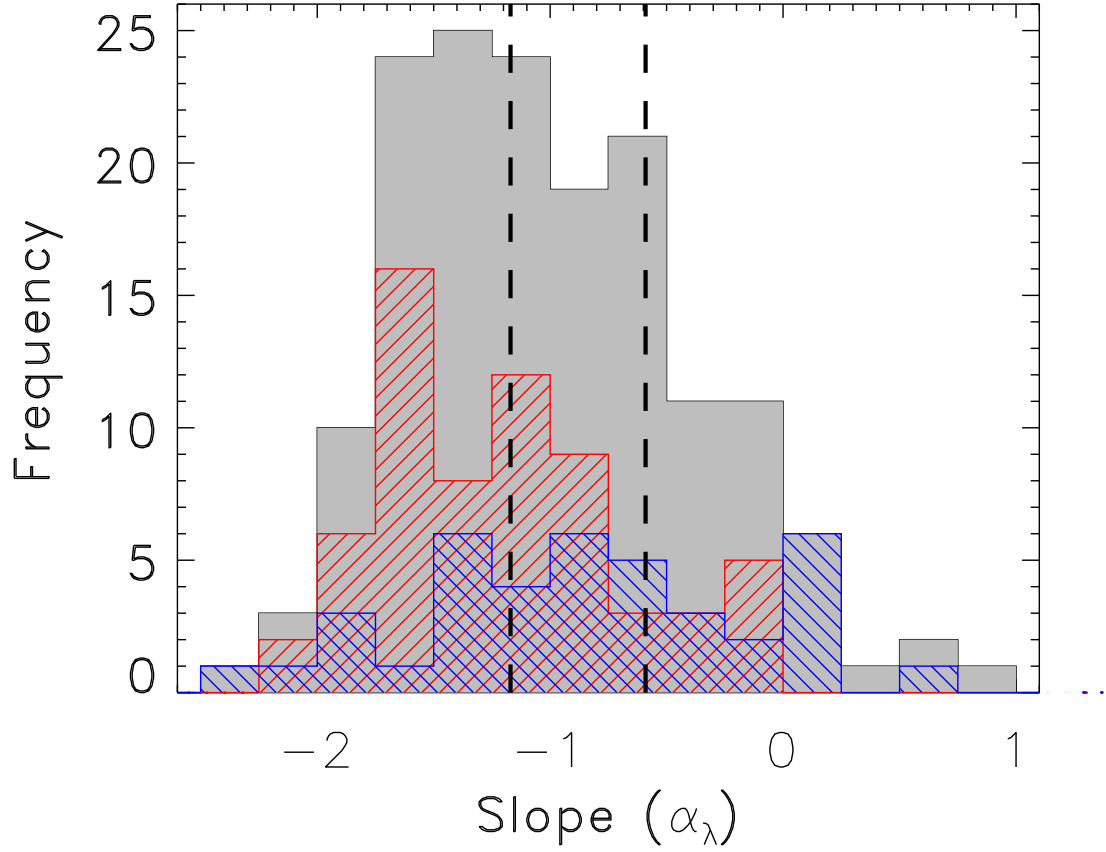} 
\caption{Distributions in spectral index, $\alpha_{\lambda}$, vs.\ redshift for 159 \HST/COS AGN,
where $(\alpha_{\lambda} + \alpha_{\nu}) = -2$.  Two vertical dashed lines mark mean values:
$\langle \alpha_{\lambda} \rangle = -0.59$ for rest-frame EUV (500-1000~\AA) and 
$\langle \alpha_{\lambda} \rangle = -1.17$  for rest-frame FUV (1200-2000~\AA).
Red histogram shows spectra using two continuum windows redward of  the 1000~\AA\ break;
blue histogram shows spectra using two continuum windows blueward of the break.  
} 
\end{figure}



\begin{figure}
\epsscale{1.0}
\plottwo{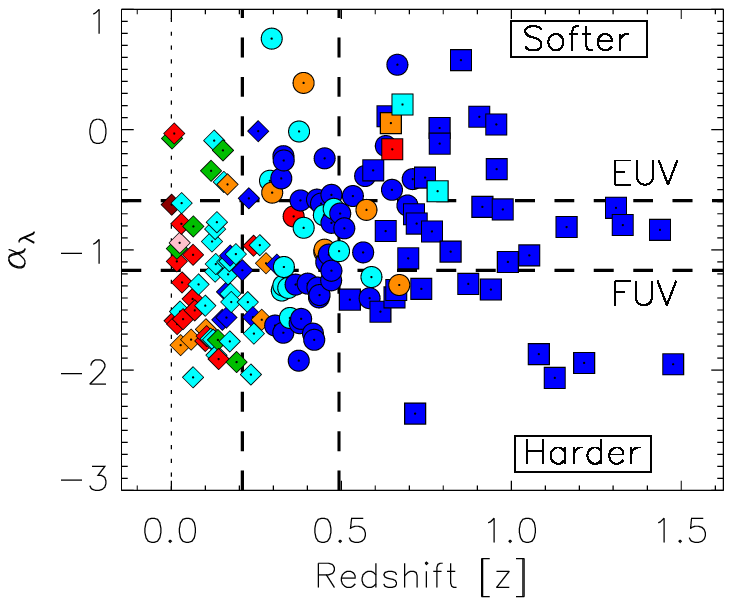}{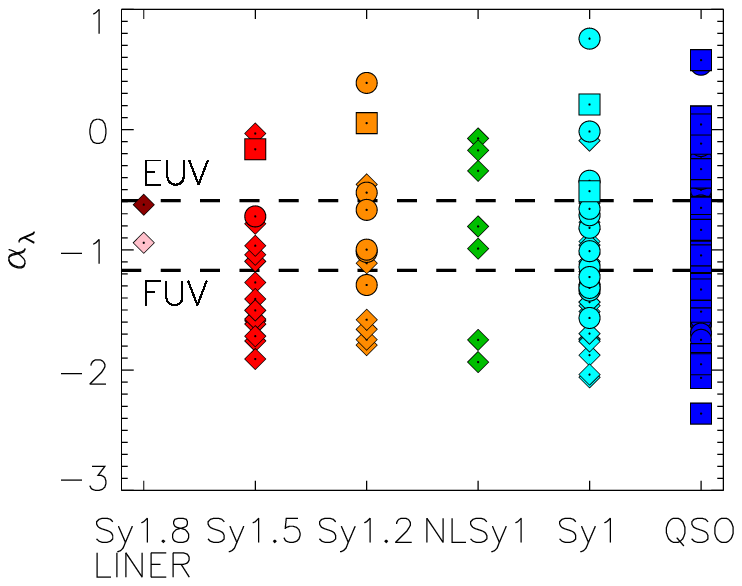}
\plottwo{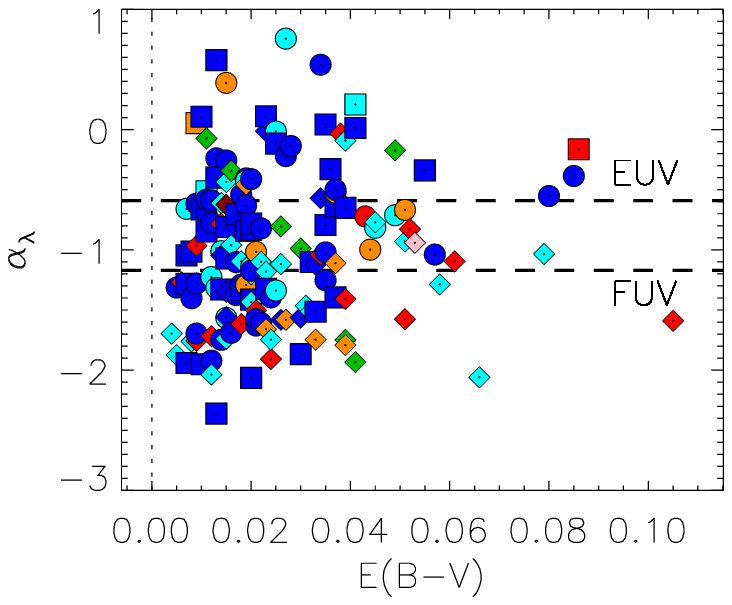}{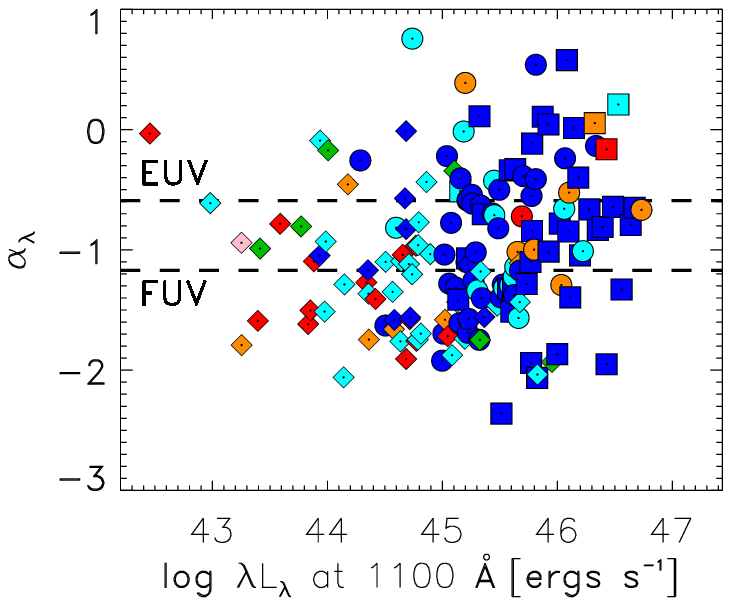}
\caption{\small
(Top Left)  Spectral index, $\alpha_{\lambda}$, vs.\ redshift for 159 \HST/COS AGN. 
(Top Right)  Index vs.\ activity type compiled by NED. (Bottom Left) Index vs. $E(B-V)$.
(Bottom Right) Index vs.\ 1100~\AA\ monochromatic luminosity.  Two low-luminosity AGN 
(NGC~4395 and NGC~4051) with $\log \lambda L_{\lambda} \leq 41$) are omitted for clarity.   
Redshift and luminosity are correlated in these composites, because of curvature in the 
underlying spectral shape.   Two horizontal lines mark mean values of this index:
$\langle \alpha_{\lambda} \rangle = -0.59$ for rest-frame EUV (500-1000~\AA); and 
$\langle \alpha_{\lambda} \rangle = -1.17$  for rest-frame FUV (1200-2000~\AA).
Vertical lines mark boundaries between targets designated as low-redshift ($z <  0.2$), 
intermediate-redshift ($0.2 < z < 0.5)$, and high-redshift ($z > 0.5$).  Colors denote AGN 
activity type: dark blue (QSO); cyan (Sy1);  green (NLSy1); orange (Sy1.2); red (Sy1.5); 
pink (LINER); brown (Sy1.8). Symbol shapes refer to windows used to fit continua:  
diamonds longward of 1000~\AA, squares shortward of 1000~\AA, circles straddling 
1000~\AA\ with one window each in EUV and FUV.
} 
\end{figure}


\clearpage


\LongTables
\begin{deluxetable*}{llllllll}
\tabletypesize{\footnotesize}
\tablecaption{COS Observations (159 AGN Targets ordered by redshift) }
\tablecolumns{8}
\tablewidth{0pt}
\tablehead{
  \colhead{AGN\tablenotemark{a}}                                    &
  \colhead{AGN\tablenotemark{a}}                                    &
  \colhead{$z$\tablenotemark{a}}                                    &
  \colhead{$\alpha_{\lambda}$}                                      &
  \colhead{$F_0$\tablenotemark{a}}                                  &       
  \colhead{$F_{\lambda}$\tablenotemark{a}}                          &
  \colhead{$\log (\lambda L_{\lambda})$\tablenotemark{a}}           &  
  \colhead{(S/N)$_{\rm res}$\tablenotemark{a}}  \\
  \colhead{Target} & \colhead{Type}    &   &   &   \colhead{(1100~\AA)}  &   \colhead{(1300~\AA)}  & (1100~\AA)  &   
            }
 \startdata

NGC4395                   & Sy~1.8     & 0.001064  & $-0.62  \pm 0.006    $ & $0.188  $ & $0.16 $ & 39.70 & 4 , 4   \\
NGC4051                   & NLSy~1   & 0.002336  & $-0.07  \pm 0.002    $ & $0.809  $ & $0.83 $ & 41.02 & 16, 13  \\
NGC3516                  & Sy~1.5     & 0.008836  & $-0.03  \pm 0.007    $ & $1.54   $ & $1.1  $ & 42.46 & 14, 12  \\
NGC3783                  & Sy~1.5     & 0.00973    & $-1.59  \pm 0.1      $ & $11.0   $ & $3.6  $ & 43.40 & 21, 17  \\
NGC7469                    & Sy~1.5    & 0.016317  & $-1.10  \pm 0.03     $ & $11.9   $ & $6.4  $ & 43.88 & 28, 18  \\
Mrk1044                      & NLSy~1   & 0.016451  & $-0.99  \pm 0.007    $ & $3.98   $ & $2.6  $ & 43.41 & 19, 13  \\
NGC5548                    & Sy~1.5     & 0.017175  & $-1.62  \pm 0.003    $ & $9.56   $ & $7.0  $ & 43.83 & 28, 19  \\
AKN564                      & LINER     & 0.024684  & $-0.94  \pm 0.02     $ & $1.20   $ & $0.67 $ & 43.25 & 9 , 7   \\
Mrk335                         & Sy~1        & 0.025785  & $-1.51  \pm 0.009    $ & $5.82   $ & $4.1  $ & 43.98 & 24, 18  \\
ESO031--G008        & Sy~1.2     & 0.027619  & $-1.79  \pm 0.01     $ & $0.958  $ & $0.62 $ & 43.25 & 9 , 7   \\
Mrk290                         & Sy~1.5     & 0.029577  & $-0.78  \pm 0.003    $ & $1.80   $ & $1.8  $ & 43.59 & 20, 16  \\
Mrk279                         & Sy~1        & 0.030451  & $-0.61  \pm 0.003    $ & $0.419  $ & $0.49 $ & 42.98 & 8 , 7   \\
Mrk817                         & Sy~1.5     & 0.031455  & $-1.27  \pm 0.0003   $ & $8.92   $ & $9.4  $ & 44.34 & 38, 24  \\
Mrk509                         & Sy~1.5     & 0.034397  & $-1.58  \pm 0.02     $ & $17.8   $ & $12.6 $ & 44.72 & 63, 55  \\
PG1011-040               & Sy~1.2     & 0.058314  & $-1.75  \pm 0.01     $ & $2.63   $ & $2.7  $ & 44.36 & 26, 13  \\
Mrk1513                       & Sy~1.5    & 0.062977  & $-1.41  \pm 0.02     $ & $2.55   $ & $5.7  $ & 44.42 & 27, 15  \\
MR2251-178               & Sy~1.5    & 0.06398     & $-1.04  \pm 0.009    $ & $4.26   $ & $25.3 $ & 44.65 & 33, 25  \\
RXJ0503.1-6634        & Sy~1       & 0.064          & $-2.06  \pm 0.05     $ & $1.30   $ & $2.2  $ & 44.14 & 16, 8   \\
SDSSJ145108.76+270926.9       & NLSy~1     & 0.0645  & $-0.81  \pm 0.008    $ & $0.539  $ & $0.92 $ & 43.77 & 10, 8   \\
RBS563                        & Sy~1.5    & 0.069         & $-1.50  \pm 0.005    $ & $0.574  $ & $9.0  $ & 43.85 & 13, 7   \\
SDSSJ031027.82-004950.7 & Sy~1 & 0.080139  & $-1.29  \pm 0.03     $ & $0.827  $ & $1.0  $ & 44.15 & 10, 8   \\
PG0804+761               & Sy~1       & 0.101         & $-1.46  \pm 0.007    $ & $11.0   $ & $9.6  $ & 45.48 & 51, 33  \\
IRAS-F22456-5125    & Sy~1.5    & 0.101         & $-1.76  \pm 0.0006   $ & $2.18   $ & $2.5  $ & 44.77 & 44, 24  \\
UKS-0242-724            & Sy~1.2    & 0.1018       & $-1.66  \pm 0.007    $ & $1.34   $ & $1.1  $ & 44.58 & 13, 9   \\
IRAS-F04250-5718    & Sy~1.5    & 0.104         & $-1.72  \pm 0.001    $ & $3.76   $ & $3.7  $ & 45.05 & 63, 34  \\
TonS210                       & Sy~1       & 0.116         & $-1.74  \pm 0.002    $ & $4.18   $ & $3.7  $ & 45.19 & 36, 21  \\
Q1230+0115                & NLSy~1  & 0.117        & $-0.34  \pm 0.002    $ & $3.33   $ & $4.2  $ & 45.10 & 51, 31  \\
HS0033+4300             & Sy~1       & 0.12           & $-0.93  \pm 0.02     $ & $0.241  $ & $0.12 $ & 43.99 & 6 , 7   \\
Mrk106                          & Sy~1       & 0.122951  & $-1.75  \pm 0.004    $ & $1.44   $ & $1.1  $ & 44.79 & 23, 14  \\
SDSSJ152139.66+033729.2   & Sy~1    & 0.126354  & $-0.09  \pm 0.01     $ & $0.192  $ & $0.17 $ & 43.94 & 5 , 4   \\
Mrk876                          & Sy~1    & 0.129            & $-1.18  \pm 0.004    $ & $4.56   $ & $3.8  $ & 45.33 & 56, 31  \\
IRASL06229-6434      & Sy1     & 0.129             & $-0.83  \pm 0.02     $ & $1.03   $ & $0.70 $ & 44.68 & 16, 9   \\
PG0838+770                & Sy~1    & 0.131           & $-1.12  \pm 0.005    $ & $1.05   $ & $0.83 $ & 44.71 & 24, 11  \\
PG1626+554                & Sy~1    & 0.133           & $-1.88  \pm 0.0005   $ & $2.41   $ & $2.2  $ & 45.08 & 24, 14  \\
QSO0045+3926           & Sy~1    & 0.134           & $-0.77  \pm 0.02     $ & $1.22   $ & $0.77 $ & 44.79 & 29, 20  \\
PKS0558-504               & NLSy~1  & 0.1372     & $-1.75  \pm 0.02     $ & $3.94   $ & $2.8  $ & 45.33 & 15, 7   \\
SDSSJ094733.21+100508.7   & Sy~1.5           & 0.139297  & $-1.91  \pm 0.004    $ & $0.868  $ & $0.65 $ & 44.68 & 11, 6   \\
SDSSJ135712.61+170444.1   & QSO               & 0.1505      & $-1.58  \pm 0.005    $ & $0.584  $ & $0.44 $ & 44.58 & 12, 8   \\
SDSSJ112114.22+032546.7   & NLSy~1         & 0.152033  & $-0.17  \pm 0.02     $ & $0.15   $ & $0.10 $ & 44.01 & 4 , 3   \\
PG1115+407                & Sy~1    & 0.154567    & $-0.44  \pm 0.002    $ & $1.04   $ & $1.0  $ & 44.86 & 19, 11  \\
SDSSJ095915.65+050355.1   & QSO               & 0.162296  & $-1.56  \pm 0.006    $ & $0.674  $ & $0.56 $ & 44.72 & 11, 7   \\
SDSSJ135625.55+251523.7   & Sy1                 & 0.164009  & $-1.35  \pm 0.004    $ & $0.464  $ & $0.39 $ & 44.57 & 8 , 5   \\
SDSSJ015530.02-085704.0   & Sy~1                & 0.164427  & $-1.10  \pm 0.004    $ & $0.4    $ & $0.35 $ & 44.51 & 9 , 6   \\
PG1202+281                & Sy~1.2                           & 0.1653    & $-0.46  \pm 0.005    $ & $0.186  $ & $0.16 $ & 44.18 & 7 , 5   \\
PG1048+342                & Sy~1                              & 0.167132  & $-1.20  \pm 0.003    $ & $0.657  $ & $0.59 $ & 44.74 & 18, 12  \\
SDSSJ121114.56+365739.5   & Sy~1               & 0.170796  & $-1.09  \pm 0.003    $ & $0.502  $ & $0.44 $ & 44.64 & 10, 7   \\
SDSSJ134231.22+382903.4   & Sy~1               & 0.171869  & $-1.76  \pm 0.002    $ & $0.488  $ & $0.49 $ & 44.63 & 10, 6   \\
1SAXJ1032.3+5051     & QSO                             & 0.173128  & $-1.05  \pm 0.007    $ & $0.095  $ & $0.08 $ & 43.93 & 9 , 5   \\
SDSSJ021218.32-073719.8   & Sy~1                & 0.17392   & $-1.37  \pm 0.006    $ & $0.251  $ & $0.21 $ & 44.36 & 8 , 5   \\
PG1116+215                 & Sy~1                             & 0.1763    & $-1.43  \pm 0.003    $ & $5.04   $ & $4.4  $ & 45.67 & 34, 22  \\
2MASX-J01013113+4229356   & Sy~1             & 0.19      & $-1.04  \pm 0.05     $ & $0.71   $ & $0.37 $ & 44.89 & 8 , 6   \\
PHL1811                        & NLSy~1                        & 0.192     & $-1.93  \pm 0.02     $ & $7.95   $ & $5.5  $ & 45.95 & 34, 18  \\
SDSSJ123604.02+264135.9   & QSO                & 0.208995  & $-1.17  \pm 0.01     $ & $0.165  $ & $0.19 $ & 44.35 & 7 , 5   \\
PG1121+422                & Sy~1                              & 0.225025  & $-1.43  \pm 0.005    $ & $0.831  $ & $0.88 $ & 45.13 & 16, 10  \\
SDSSJ001224.01-102226.5   & QSO                 & 0.228191  & $-0.57  \pm 0.006    $ & $0.283  $ & $0.21 $ & 44.67 & 7 , 5   \\
PG0953+414                & Sy~1                              & 0.2341    & $-2.04  \pm 0.001    $ & $3.80   $ & $4.3  $ & 45.83 & 33, 20  \\
SDSSJ092909.79+464424.0   & QSO                & 0.239959  & $-1.56  \pm 0.002    $ & $1.24   $ & $1.3  $ & 45.37 & 14, 9   \\
SDSSJ133053.27+311930.5   & Sy~1.5            & 0.242204  & $-0.97  \pm 0.002    $ & $0.308  $ & $0.44 $ & 44.77 & 10, 7   \\
RXJ0439.6-5311          & Sy~1                              & 0.243     & $-1.70  \pm 0.002    $ & $0.337  $ & $0.36 $ & 44.81 & 14, 8   \\
FBQSJ1010+3003       & QSO                               & 0.255778  & $-0.01  \pm 0.004    $ & $0.224  $ & $0.38 $ & 44.68 & 15, 9   \\
SDSSJ115758.72-002220.8   & Sy~1                 & 0.260247  & $-0.96  \pm 0.02     $ & $0.274  $ & $0.57 $ & 44.79 & 8 , 5   \\
SDSSJ134206.56+050523.8   & Sy~1.2             & 0.266015  & $-1.58  \pm 0.007    $ & $0.446  $ & $0.81 $ & 45.02 & 10, 7   \\
PKS1302-102               & Sy~1.2                            & 0.2784    & $-1.11  \pm 0.01     $ & $1.79   $ & $1.6  $ & 45.67 & 23, 16  \\
Ton580                           & Sy~1                               & 0.290237  & $-0.43  \pm 0.002    $ & $0.974  $ & $0.97 $ & 45.45 & 18, 12  \\
SDSSJ092837.98+602521.0   & Sy~1                 & 0.29545   & $+0.76   \pm 0.005    $ & $0.182  $ & $0.21 $ & 44.74 & 5 , 5   \\
H1821+643                   & Sy~1.2                            & 0.2968    & $-0.52  \pm 0.007    $ & $4.16   $ & $3.4  $ & 46.10 & 52, 7   \\
SDSSJ091235.42+295725.4   & QSO                 & 0.305331  & $-1.63  \pm 0.004    $ & $0.098  $ & $0.13 $ & 44.50 & 6 , 5   \\
SDSSJ082633.51+074248.3   & QSO                 & 0.310643  & $-1.12  \pm 0.009    $ & $0.491  $ & $0.52 $ & 45.22 & 9 , 6   \\
SDSSJ120720.99+262429.1   & QSO                 & 0.323529  & $-0.41  \pm 0.003    $ & $0.387  $ & $0.40 $ & 45.16 & 8 , 7   \\
SDSSJ134251.60-005345.3   & Sy~1                 & 0.325     & $-1.34  \pm 0.005    $ & $0.535  $ & $0.62 $ & 45.30 & 10, 7   \\
SDSSJ092554.43+453544.4   & QSO                 & 0.329478  & $-1.69  \pm 0.008    $ & $0.429  $ & $0.63 $ & 45.22 & 13, 11  \\
PG1001+291                & Sy~1                               & 0.3297    & $-1.30  \pm 0.003    $ & $0.999  $ & $1.1  $ & 45.59 & 18, 14  \\
PG0832+251                & QSO                                & 0.329773  & $-0.22  \pm 0.005    $ & $0.282  $ & $0.30 $ & 45.04 & 12, 10  \\
SDSSJ132704.13+443505.0   & QSO                 & 0.330709  & $-0.26  \pm 0.004    $ & $0.0494 $ & $0.07 $ & 44.29 & 4 , 4   \\
PG1216+069                & Sy~1                                & 0.3313    & $-1.14  \pm 0.02     $ & $1.10   $ & $1.2  $ & 45.63 & 20, 15  \\
RXJ2154.1-4414            & Sy~1                             & 0.344     & $-1.32  \pm 0.001    $ & $0.802  $ & $1.0  $ & 45.54 & 24, 16  \\
B0117-2837                & Sy~1                                 & 0.348858  & $-1.57  \pm 0.03     $ & $1.04   $ & $1.2  $ & 45.66 & 21, 17  \\
PG1049-005                & Sy~1.5                             & 0.3599    & $-0.72  \pm 0.02     $ & $1.03   $ & $0.82 $ & 45.69 & 11, 10  \\
SDSSJ094952.91+390203.9   & QSO                 & 0.365562  & $-1.29  \pm 0.002    $ & $0.677  $ & $0.76 $ & 45.53 & 11, 9   \\
SDSSJ132222.68+464535.2   & QSO                 & 0.374861  & $-1.92  \pm 0.002    $ & $0.189  $ & $0.25 $ & 45.00 & 8 , 6   \\
SDSSJ122035.10+385316.4   & Sy~1                & 0.375767  & $-0.02  \pm 0.01     $ & $0.29   $ & $0.28 $ & 45.18 & 6 , 6   \\
SDSSJ024250.85-075914.2   & QSO                 & 0.377651  & $-1.61  \pm 0.003    $ & $0.263  $ & $0.33 $ & 45.15 & 7 , 6   \\
SDSSJ123335.07+475800.4   & QSO                & 0.38223    & $-0.59  \pm 0.009    $ & $0.304  $ & $0.30 $ & 45.22 & 9 , 8   \\
SDSSJ134246.89+184443.6   & QSO                & 0.382     & $-1.57  \pm 0.004    $ & $0.308  $ & $0.40 $ & 45.23 & 8 , 7   \\
SDSSJ121037.56+315706.0   & Sy~1.2            & 0.389041  & $+0.39   \pm 0.003    $ & $0.276  $ & $0.26 $ & 45.20 & 7 , 8   \\
HB89-0202-765           & Sy~1                              & 0.38939   & $-0.82  \pm 0.01     $ & $0.0688 $ & $0.23 $ & 44.60 & 9 , 8   \\
SDSSJ110312.93+414154.9   & QSO                & 0.401023  & $-1.28  \pm 0.009    $ & $0.186  $ & $0.21 $ & 45.06 & 9 , 7   \\
SDSSJ133045.15+281321.4   & QSO                & 0.416754  & $-1.70  \pm 0.006    $ & $0.151  $ & $0.21 $ & 45.01 & 10, 7   \\
SDSSJ111754.31+263416.6   & QSO                & 0.420466  & $-1.75  \pm 0.03     $ & $0.303  $ & $0.38 $ & 45.32 & 9 , 7   \\
SDSSJ143511.53+360437.2   & QSO                & 0.428593  & $-0.58  \pm 0.3      $ & $0.231  $ & $0.24 $ & 45.22 & 7 , 6   \\
HE0435-5304               & QSO                               & 0.427      & $-1.31  \pm 0.0009   $ & $0.179  $ & $0.24 $ & 45.12 & 12, 7   \\
SDSSJ110406.94+314111.4   & QSO                & 0.434356  & $-1.39  \pm 0.005    $ & $0.432  $ & $0.52 $ & 45.51 & 10, 8   \\
B0120-28                       & QSO                               & 0.436018  & $-1.38  \pm 0.002    $ & $0.558  $ & $0.54 $ & 45.62 & 15, 12  \\
SDSSJ161649.42+415416.3   & QSO                 & 0.440417  & $-0.62  \pm 0.003    $ & $0.236  $ & $0.23 $ & 45.26 & 8 , 6   \\
SDSSJ080359.23+433258.4   & Sy~1                & 0.448706  & $-0.71  \pm 0.02     $ & $0.351  $ & $0.13 $ & 45.45 & 7 , 5   \\
TON236                          & Sy~1.2                           & 0.4473   & $-1.02  \pm 0.003    $ & $0.556  $ & $0.59 $ & 45.65 & 15, 12  \\
PG0003+158                & Sy~1.2                            & 0.4509    & $-1.00  \pm 0.05     $ & $0.769  $ & $0.70 $ & 45.80 & 20, 15  \\
HE0153-4520               & QSO                                & 0.451     & $-0.24  \pm 0.002    $ & $1.43   $ & $1.4  $ & 46.07 & 22, 15  \\
SDSSJ100902.06+071343.8   & QSO                 & 0.455631  & $-1.10  \pm 0.1      $ & $0.194  $ & $0.16 $ & 45.21 & 6 , 5   \\
SDSSJ091029.75+101413.6   & QSO                 & 0.463194  & $-1.04  \pm 4.4      $ & $0.12   $ & $0.07 $ & 45.02 & 4 , 5   \\
SDSSJ082024.21+233450.4   & QSO                 & 0.470212  & $-1.25  \pm 0.01     $ & $0.207  $ & $0.21 $ & 45.27 & 9 , 5   \\
SDSSJ161916.54+334238.4   & QSO                 & 0.470946  & $-1.17  \pm 0.9      $ & $0.523  $ & $0.40 $ & 45.67 & 12, 11  \\
SDSSJ092554.70+400414.1   & QSO                 & 0.471139  & $-0.78  \pm 0.007    $ & $0.131  $ & $0.14 $ & 45.08 & 6 , 4   \\
SDSSJ123304.05-003134.1   & QSO                  & 0.471167  & $-0.54  \pm 0.007    $ & $0.199  $ & $0.22 $ & 45.26 & 9 , 6   \\
PG1259+593                & Sy~1                               & 0.4778    & $-0.66  \pm 0.001    $ & $1.22   $ & $1.4  $ & 46.06 & 29, 19  \\
HE0226-4110               & Sy~1                               & 0.493368  & $-1.01  \pm 0.003    $ & $1.65   $ & $2.1  $ & 46.22 & 29, 19  \\
SDSSJ155048.29+400144.9   & QSO                 & 0.496843  & $-0.70  \pm 0.2      $ & $0.265  $ & $0.19 $ & 45.44 & 7 , 6   \\
HS1102+3441              & QSO                                & 0.508847  & $-0.82  \pm 0.008    $ & $0.281  $ & $0.30 $ & 45.49 & 15, 10  \\
SDSSJ113327.78+032719.1   & QSO                 & 0.525073  & $-1.41  \pm 0.3      $ & $0.117  $ & $0.12 $ & 45.14 & 7 , 5   \\
SDSSJ025937.46+003736.3   & QSO                 & 0.534178  & $-0.55  \pm 0.05     $ & $0.486  $ & $0.32 $ & 45.78 & 7 , 7   \\
SDSSJ094331.61+053131.4   & QSO                 & 0.564336  & $-1.02  \pm 0.1      $ & $0.139  $ & $0.17 $ & 45.29 & 6 , 4   \\
SDSSJ040148.98-054056.5   & QSO                  & 0.570076  & $-0.39  \pm 0.03     $ & $0.35   $ & $0.16 $ & 45.70 & 8 , 6   \\
PKS0405-123               & Sy~1.2                            & 0.574     & $-0.67  \pm 0.01     $ & $3.69   $ & $3.2  $ & 46.73 & 63, 29  \\
SDSSJ124154.02+572107.3   & QSO                 & 0.583237  & $-1.40  \pm 0.003    $ & $0.145  $ & $0.22 $ & 45.34 & 9 , 6   \\
SDSSJ095000.73+483129.3   & Sy~1                & 0.588734  & $-1.23  \pm 0.004    $ & $0.268  $ & $0.42 $ & 45.62 & 8 , 5   \\
SDSSJ225738.20+134045.4   & QSO                 & 0.593937  & $-0.34  \pm 0.04     $ & $0.243  $ & $0.19 $ & 45.59 & 7 , 6   \\
SDSSJ022614.46+001529.7   & QSO                 & 0.6151    & $-1.51  \pm 0.01     $ & $0.229  $ & $0.34 $ & 45.60 & 10, 6   \\
SDSSJ105945.23+144142.9   & QSO                 & 0.630543  & $-0.84  \pm 0.01     $ & $0.327  $ & $0.29 $ & 45.78 & 10, 8   \\
HE0238-1904               & QSO                               & 0.631     & $-0.14  \pm 0.005    $ & $1.17   $ & $1.3  $ & 46.34 & 23, 18  \\
SDSSJ111239.11+353928.2   & QSO                & 0.635784  & $+0.11   \pm 0.01     $ & $0.112  $ & $0.11 $ & 45.32 & 6 , 4   \\
3C263                              & Sy~1.2                         & 0.646     & $+0.05  \pm 0.02     $ & $1.08   $ & $0.87 $ & 46.33 & 30, 20  \\
SDSSJ093518.19+020415.5   & QSO                & 0.649117  & $-0.50  \pm 0.01     $ & $0.158  $ & $0.17 $ & 45.49 & 6 , 5   \\
PKS0637-752               & Sy~1.5                           & 0.653      & $-0.16  \pm 0.1      $ & $1.35   $ & $0.70 $ & 46.43 & 20, 15  \\
SDSSJ080908.13+461925.6   & QSO                & 0.656338  & $-1.39  \pm 0.02     $ & $0.637  $ & $0.79 $ & 46.11 & 12, 10  \\
SDSSJ105958.82+251708.8   & QSO                & 0.661907  & $-1.33  \pm 0.004    $ & $0.198  $ & $0.33 $ & 45.61 & 9 , 7   \\
SDSSJ154553.48+093620.5   & QSO                & 0.665     & $+0.54   \pm 1.2      $ & $0.31   $ & $-0.0 $ & 45.81 & 1 , 6   \\
3C57                      & Sy~1.2                                   & 0.670527  & $-1.29  \pm 0.04     $ & $0.506  $ & $0.77 $ & 46.03 & 21, 12  \\
PKS0552-640               & Sy~1                              & 0.68      & $+0.21   \pm 0.2      $ & $1.53   $ & $1.0  $ & 46.53 & 24, 18  \\
SDSSJ151428.64+361957.9   & QSO                & 0.694596  & $-0.63  \pm 0.02     $ & $0.0922 $ & $0.14 $ & 45.33 & 4 , 4   \\
SDSSJ144511.28+342825.4   & QSO                & 0.696951  & $-1.07  \pm 0.008    $ & $0.0697 $ & $0.11 $ & 45.21 & 7 , 5   \\
SDSSJ113457.62+255527.9   & QSO                & 0.710078  & $-0.41  \pm 0.06     $ & $0.264  $ & $0.24 $ & 45.81 & 8 , 7   \\
SDSSJ155504.39+362848.0   & QSO                & 0.713654  & $-0.70  \pm 0.01     $ & $0.0901 $ & $0.16 $ & 45.35 & 7 , 4   \\
SDSSJ124511.25+335610.1   & QSO                & 0.717     & $-2.36  \pm 0.8      $ & $0.129  $ & $0.23 $ & 45.51 & 7 , 8   \\
SDSSJ155304.92+354828.6   & QSO                & 0.721814  & $-0.78  \pm 0.01     $ & $0.409  $ & $0.50 $ & 46.02 & 8 , 6   \\
SDSSJ091440.38+282330.6   & QSO                & 0.735345  & $-1.32  \pm 0.01     $ & $0.139  $ & $0.23 $ & 45.57 & 8 , 6   \\
SDSSJ100102.55+594414.3   & QSO                & 0.746236  & $-0.40  \pm 0.3      $ & $0.546  $ & $0.58 $ & 46.18 & 11, 9   \\
SBS1108+560               & QSO                              & 0.766619  & $-0.85  \pm 0.7      $ & $0.419  $ & $0.01 $ & 46.10 & 4 , 10  \\
SDSSJ143726.14+504555.8   & Sy~1                & 0.783319  & $-0.51  \pm 0.1      $ & $0.0456 $ & $0.04 $ & 45.16 & 4 , 4   \\
SDSSJ102218.99+013218.8   & QSO                & 0.789304  & $+0.01   \pm 0.03     $ & $0.437  $ & $0.34 $ & 46.15 & 8 , 6   \\
SDSSJ234500.43-005936.0   & QSO                 & 0.789429  & $-0.12  \pm 0.08     $ & $0.188  $ & $0.12 $ & 45.78 & 7 , 5   \\
SDSSJ101622.60+470643.3   & QSO                & 0.821527  & $-1.01  \pm 0.4      $ & $0.238  $ & $0.40 $ & 45.92 & 8 , 5   \\
SBS1122+594               & QSO                              & 0.852     & $+0.58   \pm 0.1      $ & $0.313  $ & $0.26 $ & 46.08 & 12, 9   \\
SDSSJ141910.20+420746.9   & QSO                & 0.873501  & $-1.28  \pm 0.07     $ & $0.132  $ & $0.21 $ & 45.73 & 7 , 4   \\
SDSSJ112244.89+575543.0   & QSO                & 0.905906  & $+0.11   \pm 0.005    $ & $0.166  $ & $0.20 $ & 45.87 & 7 , 5   \\
FBQSJ0751+2919            & QSO                          & 0.915     & $-0.64  \pm 0.02     $ & $0.66   $ & $0.68 $ & 46.48 & 26, 19  \\
PG1407+265                & QSO                               & 0.946      & $-1.33  \pm 0.06     $ & $0.733  $ & $1.1  $ & 46.56 & 33, 20  \\
HB89-0107-025-NED05       & QSO                    & 0.956     & $+0.04   \pm 0.07     $ & $0.162  $ & $0.12 $ & 45.92 & 12, 8   \\
LBQS-0107-0235            & QSO                           & 0.957039  & $-0.33  \pm 0.08     $ & $0.0834 $ & $0.07 $ & 45.63 & 11, 10  \\
PG1148+549                & QSO                               & 0.975     & $-0.66  \pm 0.01     $ & $0.351  $ & $0.46 $ & 46.28 & 25, 15  \\
SDSSJ084349.49+411741.6   & QSO                & 0.989986  & $-1.10  \pm 0.2      $ & $0.104  $ & $0.08 $ & 45.77 & 5 , 5   \\
HE0439-5254               & QSO                               & 1.053     & $-1.05  \pm 0.04     $ & $0.241  $ & $0.38 $ & 46.20 & 14, 9   \\
SDSS-J100535.24+013445.7  & QSO                & 1.0809    & $-1.87  \pm 0.09     $ & $0.143  $ & $0.27 $ & 46.00 & 11, 11  \\
FIRST-J020930.7-043826    & QSO                    & 1.131     & $-2.06  \pm 0.02     $ & $0.0858 $ & $0.25 $ & 45.82 & 13, 13  \\
PG1206+459                & QSO                               & 1.16254   & $-0.81  \pm 1.1      $ & $0.296  $ & $0.40 $ & 46.39 & 21, 16  \\
PG1338+416                & QSO                               & 1.21422   & $-1.94  \pm 0.4      $ & $0.0632 $ & $0.18 $ & 45.77 & 16, 11  \\
LBQS-1435-0134            & QSO                           & 1.30791   & $-0.65  \pm 0.01     $ & $0.418  $ & $0.59 $ & 46.67 & 26, 20  \\
PG1522+101               & QSO                              & 1.32785   & $-0.79  \pm 0.3      $ & $0.372  $ & $0.43 $ & 46.64 & 19, 14  \\
Q0232-042                 & QSO                                 & 1.43737   & $-0.83  \pm 0.3      $ & $0.158  $ & $0.22 $ & 46.35 & 15, 12  \\
PG1630+377                & QSO                              & 1.47607   & $-1.95  \pm 0.1      $ & $0.176  $ & $0.70 $ & 46.43 & 25, 10  

\enddata

\tablenotetext{a}{Our 159 AGN targets, types, redshifts, fluxes, spectral indices, luminosities, and S/N ratios.  
   All fluxes in units of $10^{-14}$~erg~cm$^{-2}$~s$^{-1}$~{\AA}$^{-1}$. Rest-frame, dereddened spectral 
   distributions are fitted to power laws, $F_{\lambda} = F_0 (\lambda / 1100~{\rm \AA})^{\alpha_{\lambda}}$.   
   Wavelength index $\alpha_{\lambda}$ corresponds to frequency index $\alpha_{\nu} = -[2 + \alpha_{\lambda}]$.   
   The eight columns show:  
   (1)  AGN target;  (2) AGN type;  (3) AGN redshift;  (4) Fitted spectral index, $\alpha_{\lambda}$, with 
    statistical errors;   
   (5) Rest-frame flux normalization $F_0$ at 1100~\AA; (6) Observed flux $F_{\lambda}$ at 1300~\AA;
   (7) Band luminosity, $\lambda L_{\lambda}$ at 1100~\AA\ (in erg~s$^{-1}$); 
   (8) signal-to-noise at 1250~\AA\ and 1550~\AA\ for data with G130M (1132--1460~\AA) and 
   G160M (1394--1798~\AA) gratings, respectively.  
   Flux at 1300~\AA\  for SBS~1108+560 (noted with $^*$) is low,  owing to LyC absorption 
   ($\lambda < 1334$~\AA) from a LLS at $z = 0.46335$.   Finn \etal\ (2014) fitted a harder spectral index,
   $\alpha_{\lambda} = -0.64$, for J0209-0438 at $z = 1.131$, using additional COS/G230L spectra extending to 
   longer wavelengths.  }
 
\end{deluxetable*}


\newpage


\begin{deluxetable}{lllc}
\tabletypesize{\footnotesize}
\tablecaption{Lyman-limit systems and partial Lyman-limit systems}
\tablecolumns{4}
\tablewidth{0pt}
\tablehead{
\colhead{AGN Target}   &   \colhead{$z_{\rm LLS}$\tablenotemark{a}}   &   \colhead{log~N$_{\rm HI}$\tablenotemark{a}}   
    & \colhead{$b$\tablenotemark{a} }  \\
     &  \colhead{(redshift)}    &   \colhead{(\NHI\ in cm$^{-2}$)}    &   \colhead{ (km~s$^{-1}$) }  
 }
\startdata
SDSSJ115758.72-002220.8   & 0.25661   & $15.25 \pm 0.03     $ &  25  \\
SDSSJ092554.43+453544.4   & 0.25057   & $15.14 \pm 0.03     $ &  25  \\
                          & 0.30959   & $15.27 \pm 0.03     $ &  25  \\
PG1216+069                & 0.28231   & $16.29 \pm 0.01     $ &  34  \\
B0117-2837                & 0.34833   & $15.52 \pm 0.05     $ &  25  \\
                          & 0.34866   & $16.02 \pm 0.02     $ &  25  \\
SDSSJ122035.10+385316.4   & 0.27332   & $15.61 \pm 0.02     $ &  25  \\
SDSSJ123335.07+475800.4   & 0.28495   & $15.41 \pm 0.03     $ &  25  \\
HB89-0202-765             & 0.30657   & $14.95 \pm 0.03     $ &  25  \\
SDSSJ110312.93+414154.9   & 0.27116   & $15.11 \pm 0.03     $ &  25  \\
SDSSJ133045.15+281321.4   & 0.27553   & $15.20 \pm 0.02     $ &  25  \\
SDSSJ111754.31+263416.6   & 0.35193   & $16.14 \pm 0.02     $ &  25  \\
SDSSJ143511.53+360437.2   & 0.26246   & $15.24 \pm 0.02     $ &  25  \\
                          & 0.37292   & $16.72 \pm 0.06     $ &  25  \\
                          & 0.3876    & $16.15 \pm 0.02     $ &  25  \\
PG0003+158                & 0.30573   & $15.47 \pm 0.01     $ &  25  \\
                          & 0.31215   & $14.38 \pm 0.02     $ &  25  \\
                          & 0.34787   & $15.98 \pm 0.04     $ &  17  \\
                          & 0.36619   & $15.10 \pm 0.02     $ &  25  \\
                          & 0.37034   & $14.68 \pm 0.06     $ &  25  \\
                          & 0.38612   & $14.84 \pm 0.01     $ &  25  \\
                          & 0.40137   & $15.04 \pm 0.02     $ &  25  \\
                          & 0.42184   & $14.77 \pm 0.02     $ &  25  \\
HE0153-4520               & 0.40051   & $14.52 \pm 0.02     $ &  25  \\
SDSSJ100902.06+071343.8   & 0.35586   & $17.41 \pm 0.04     $ &  25  \\
                          & 0.3745    & $14.28 \pm 0.06     $ &  25  \\
                          & 0.37554   & $15.62 \pm 0.04     $ &  25  \\
                          & 0.37624   & $14.86 \pm 0.06     $ &  25  \\
                          & 0.41401   & $15.15 \pm 0.02     $ &  25  \\
SDSSJ091029.75+101413.6   & 0.2634    & $16.86 \pm 0.5      $ &  25  \\
                          & 0.26375   & $14.17 \pm 0.1      $ &  25  \\
                          & 0.26432   & $14.96 \pm 0.03     $ &  25  \\
                          & 0.41924   & $17.14 \pm 0.8      $ &  29  \\
SDSSJ082024.21+233450.4   & 0.45424   & $14.56 \pm 0.02     $ &  25  \\
SDSSJ161916.54+334238.4   & 0.2694    & $16.40 \pm 0.03     $ &  25  \\
                          & 0.26988   & $14.72 \pm 0.08     $ &  25  \\
                          & 0.27086   & $14.85 \pm 0.02     $ &  25  \\
                          & 0.2716    & $14.58 \pm 0.08     $ &  25  \\
                          & 0.42676   & $14.93 \pm 0.07     $ &  25  \\
                          & 0.44231   & $15.81 \pm 0.02     $ &  25  \\
                          & 0.47091   & $16.84 \pm 0.1      $ &  25  \\
                          & 0.47179   & $14.47 \pm 0.03     $ &  25  \\
SDSSJ092554.70+400414.1   & 0.2477    & $19.26 \pm 0.06     $ &  59  \\
                          & 0.25283   & $14.84 \pm 0.05     $ &  25  \\
SDSSJ123304.05-003134.1   & 0.31875   & $15.51 \pm 0.01     $ &  25  \\
                          & 0.43061   & $14.89 \pm 0.02     $ &  25  \\
PG1259+593                & 0.2924    & $14.49 \pm 0.02     $ &  25  \\
HE0226-4110               & 0.24525   & $14.24 \pm 0.03     $ &  25  \\
                          & 0.49252   & $14.64 \pm 0.05     $ &  25  \\
SDSSJ155048.29+400144.9   & 0.31257   & $16.62 \pm 0.06     $ &  41  \\
                          & 0.42739   & $15.64 \pm 0.02     $ &  25  \\
                          & 0.4919    & $16.57 \pm 0.02     $ &  25  \\
                          & 0.49255   & $15.69 \pm 0.05     $ &  25  \\
HS1102+3441               & 0.26164   & $14.95 \pm 0.03     $ &  25  \\
                          & 0.28916   & $14.49 \pm 0.04     $ &  25  \\
                          & 0.28986   & $14.89 \pm 0.03     $ &  25  \\
                          & 0.31039   & $15.07 \pm 0.01     $ &  25  \\
                          & 0.33246   & $14.37 \pm 0.02     $ &  25  \\
                          & 0.50607   & $15.07 \pm 0.02     $ &  25  \\
SDSSJ113327.78+032719.1   & 0.24663   & $17.53 \pm 0.1      $ &  25  \\
                          & 0.30216   & $14.44 \pm 0.04     $ &  25  \\
                          & 0.45225   & $15.22 \pm 0.03     $ &  25  \\
SDSSJ094331.61+053131.4   & 0.35464   & $16.12 \pm 0.09     $ &  91  \\
SDSSJ040148.98-054056.5   & 0.32381   & $15.37 \pm 0.01     $ &  25  \\
                          & 0.36547   & $14.60 \pm 0.05     $ &  25  \\
PKS0405-123               & 0.36077   & $15.04 \pm 0.02     $ &  25  \\
                          & 0.4057    & $14.83 \pm 0.02     $ &  25  \\
SDSSJ095000.73+483129.3   & 0.48502   & $15.07 \pm 0.03     $ &  25  \\
SDSSJ225738.20+134045.4   & 0.37712   & $15.07 \pm 0.02     $ &  25  \\
                          & 0.3787    & $14.90 \pm 0.08     $ &  25  \\
                          & 0.49905   & $15.72 \pm 0.02     $ &  25  \\
SDSSJ022614.46+001529.7   & 0.4161    & $14.77 \pm 0.02     $ &  25  \\
SDSSJ105945.23+144142.9   & 0.34074   & $15.33 \pm 0.02     $ &  25  \\
                          & 0.46567   & $15.65 \pm 0.03     $ &  25  \\
                          & 0.57638   & $15.51 \pm 0.03     $ &  25  \\
                          & 0.61727   & $14.71 \pm 0.02     $ &  25  \\
HE0238-1904               & 0.3441    & $14.69 \pm 0.02     $ &  25  \\
                          & 0.35534   & $14.87 \pm 0.03     $ &  25  \\
                          & 0.40102   & $14.91 \pm 0.02     $ &  25  \\
                          & 0.42424   & $15.03 \pm 0.02     $ &  25  \\
SDSSJ111239.11+353928.2   & 0.24679   & $15.45 \pm 0.03     $ &  25  \\
3C263                     & 0.32545   & $15.44 \pm 0.02     $ &  25  \\
                          & 0.52796   & $15.55 \pm 0.01     $ &  25  \\
SDSSJ093518.19+020415.5   & 0.35457   & $15.26 \pm 0.03     $ &  25  \\
                          & 0.42852   & $14.69 \pm 0.03     $ &  25  \\
PKS0637-752               & 0.24326   & $15.86 \pm 0.09     $ &  38  \\
                          & 0.41755   & $15.42 \pm 0.01     $ &  25  \\
                          & 0.4528    & $15.47 \pm 0.02     $ &  25  \\
                          & 0.46847   & $16.08 \pm 0.03     $ &  25  \\
SDSSJ080908.13+461925.6   & 0.61917   & $16.15 \pm 0.01     $ &  25  \\
SDSSJ154553.48+093620.5   & 0.47379   & $17.25 \pm 0.2      $ &  25  \\
                          & 0.47623   & $15.62 \pm 0.02     $ &  25  \\
3C57                      & 0.24988   & $15.67 \pm 0.009    $ &  25  \\
                          & 0.29224   & $14.84 \pm 0.02     $ &  25  \\
                          & 0.32332   & $16.29 \pm 0.01     $ &  76  \\
                          & 0.32827   & $15.53 \pm 0.02     $ &  25  \\
                          & 0.38329   & $14.75 \pm 0.02     $ &  25  \\
                          & 0.5332    & $14.22 \pm 0.08     $ &  25  \\
PKS0552-640               & 0.34517   & $16.71 \pm 0.03     $ &  25  \\
                          & 0.34592   & $14.18 \pm 0.04     $ &  25  \\
                          & 0.446     & $15.89 \pm 0.04     $ &  30  \\
                          & 0.63017   & $14.78 \pm 0.06     $ &  25  \\
SDSSJ151428.64+361957.9   & 0.41065   & $17.93 \pm 0.2      $ &  42  \\
SDSSJ144511.28+342825.4   & 0.60722   & $15.51 \pm 0.02     $ &  25  \\
SDSSJ113457.62+255527.9   & 0.2469    & $14.66 \pm 0.03     $ &  25  \\
                          & 0.43233   & $16.40 \pm 0.03     $ &  53  \\
                          & 0.50265   & $15.19 \pm 0.02     $ &  25  \\
                          & 0.66824   & $15.01 \pm 0.05     $ &  25  \\
SDSSJ155504.39+362848.0   & 0.30689   & $14.57 \pm 0.02     $ &  25  \\
                          & 0.36504   & $14.94 \pm 0.02     $ &  25  \\
                          & 0.57611   & $15.52 \pm 0.03     $ &  25  \\
                          & 0.60275   & $15.36 \pm 0.02     $ &  25  \\
SDSSJ124511.25+335610.1   & 0.31802   & $14.80 \pm 0.03     $ &  25  \\
                          & 0.44947   & $15.44 \pm 0.1      $ &  25  \\
                          & 0.55682   & $16.50 \pm 0.2      $ &  25  \\
                          & 0.58762   & $15.14 \pm 0.03     $ &  25  \\
                          & 0.63215   & $15.92 \pm 0.06     $ &  65  \\
                          & 0.64496   & $15.70 \pm 0.04     $ &  25  \\
                          & 0.64862   & $15.17 \pm 0.04     $ &  25  \\
                          & 0.68918   & $16.68 \pm 0.2      $ &  25  \\
                          & 0.71297   & $16.32 \pm 0.1      $ &  35  \\
SDSSJ155304.92+354828.6   & 0.4756    & $15.43 \pm 0.03     $ &  25  \\
                          & 0.52027   & $15.18 \pm 0.03     $ &  25  \\
SDSSJ091440.38+282330.6   & 0.24426   & $15.39 \pm 0.01     $ &  25  \\
                          & 0.59969   & $15.33 \pm 0.02     $ &  25  \\
SDSSJ100102.55+594414.3   & 0.30355   & $17.27 \pm 0.04     $ &  25  \\
                          & 0.4159    & $16.61 \pm 0.02     $ &  25  \\
SBS1108+560               & 0.28646   & $15.91 \pm 0.2      $ &  50  \\
                          & 0.46334   & $17.06 \pm 0.1      $ &  35  \\
                          & 0.61765   & $15.07 \pm 0.03     $ &  25  \\
                          & 0.68267   & $15.36 \pm 0.02     $ &  25  \\
SDSSJ143726.14+504555.8   & 0.25065   & $15.77 \pm 0.06     $ &  25  \\
                          & 0.56945   & $15.23 \pm 0.03     $ &  25  \\
                          & 0.76901   & $15.52 \pm 0.08     $ &  25  \\
                          & 0.77109   & $14.94 \pm 0.1      $ &  25  \\
                          & 0.77248   & $16.07 \pm 0.1      $ &  36  \\
SDSSJ102218.99+013218.8   & 0.39907   & $13.41 \pm 0.09     $ &  25  \\
                          & 0.7425    & $15.35 \pm 0.1      $ &  25  \\
SDSSJ234500.43-005936.0   & 0.2539    & $16.08 \pm 0.03     $ &  25  \\
                          & 0.54818   & $15.96 \pm 0.07     $ &  34  \\
SDSSJ101622.60+470643.3   & 0.4321    & $15.59 \pm 0.02     $ &  25  \\
                          & 0.66475   & $15.92 \pm 0.1      $ &  25  \\
                          & 0.72766   & $16.26 \pm 0.3      $ &  14  \\
                          & 0.74627   & $15.43 \pm 0.05     $ &  25  \\
SBS1122+594               & 0.31236   & $14.95 \pm 0.02     $ &  25  \\
                          & 0.31784   & $14.43 \pm 0.04     $ &  25  \\
                          & 0.35115   & $15.14 \pm 0.02     $ &  25  \\
                          & 0.3919    & $15.18 \pm 0.06     $ &  25  \\
                          & 0.55744   & $15.82 \pm 0.03     $ &  25  \\
                          & 0.55817   & $16.42 \pm 0.02     $ &  25  \\
                          & 0.5698    & $14.72 \pm 0.04     $ &  25  \\
                          & 0.67835   & $15.97 \pm 0.08     $ &  19  \\
SDSSJ141910.20+420746.9   & 0.289     & $16.17 \pm 0.03     $ &  25  \\
                          & 0.42561   & $16.02 \pm 0.02     $ &  25  \\
                          & 0.52221   & $15.87 \pm 0.02     $ &  25  \\
                          & 0.53461   & $16.06 \pm 0.07     $ &  25  \\
                          & 0.60842   & $15.72 \pm 0.02     $ &  25  \\
                          & 0.80463   & $15.95 \pm 0.04     $ &  25  \\
                          & 0.84523   & $16.20 \pm 0.03     $ &  25  \\
SDSSJ112244.89+575543.0   & 0.39798   & $14.78 \pm 0.03     $ &  25  \\
FBQSJ0751+2919            & 0.43187   & $15.69 \pm 0.01     $ &  25  \\
                          & 0.49455   & $15.56 \pm 0.02     $ &  25  \\
                          & 0.82902   & $15.99 \pm 0.02     $ &  25  \\
PG1407+265                & 0.29717   & $15.13 \pm 0.03     $ &  25  \\
                          & 0.32569   & $15.20 \pm 0.01     $ &  25  \\
                          & 0.57488   & $15.56 \pm 0.01     $ &  25  \\
                          & 0.59964   & $15.78 \pm 0.03     $ &  26  \\
                          & 0.68278   & $16.39 \pm 0.03     $ &  33  \\
                          & 0.81699   & $15.62 \pm 0.02     $ &  25  \\
HB89-0107-025-NED05       & 0.39909   & $16.59 \pm 0.02     $ &  30  \\
                          & 0.53546   & $15.08 \pm 0.03     $ &  25  \\
                          & 0.7178    & $15.37 \pm 0.01     $ &  25  \\
                          & 0.8093    & $15.17 \pm 0.05     $ &  25  \\
                          & 0.87569   & $15.51 \pm 0.06     $ &  25  \\
LBQS-0107-0235            & 0.53635   & $15.81 \pm 0.05     $ &  86  \\
                          & 0.71892   & $15.54 \pm 0.02     $ &  25  \\
                          & 0.87636   & $15.77 \pm 0.02     $ &  25  \\
PG1148+549                & 0.25242   & $15.26 \pm 0.02     $ &  25  \\
                          & 0.57785   & $15.24 \pm 0.01     $ &  25  \\
                          & 0.68864   & $15.52 \pm 0.03     $ &  25  \\
                          & 0.90485   & $15.34 \pm 0.05     $ &  25  \\
SDSSJ084349.49+411741.6   & 0.53258   & $16.67 \pm 0.05     $ &  25  \\
                          & 0.53346   & $16.10 \pm 0.03     $ &  25  \\
                          & 0.54106   & $15.30 \pm 0.02     $ &  25  \\
                          & 0.54353   & $15.57 \pm 0.04     $ &  25  \\
HE0439-5254               & 0.328     & $15.67 \pm 0.01     $ &  25  \\
                          & 0.61508   & $16.25 \pm 0.04     $ &  52  \\
                          & 0.86515   & $15.58 \pm 0.03     $ &  25  \\
SDSS-J100535.24+013445.7  & 0.41753   & $15.04 \pm 0.05     $ &  25  \\
                          & 0.41853   & $16.37 \pm 0.03     $ &  25  \\
                          & 0.41963   & $15.83 \pm 0.02     $ &  25  \\
                          & 0.83711   & $16.81 \pm 0.01     $ &  25  \\
                          & 0.83938   & $16.10 \pm 0.04     $ &  25  \\
FIRST-J020930.7-043826    & 0.39035   & $18.00 \pm 0.2      $ &  49  \\
                          & 0.8268    & $15.05 \pm 0.04     $ &  25  \\
PG1206+459                & 0.40852   & $15.71 \pm 0.02     $ &  25  \\
                          & 0.41412   & $15.69 \pm 0.04     $ &  25  \\
                          & 0.92772   & $17.03 \pm 0.08     $ &  46  \\
PG1338+416                & 0.34886   & $16.37 \pm 0.06     $ &  41  \\
                          & 0.46369   & $15.28 \pm 0.01     $ &  25  \\
                          & 0.62136   & $16.30 \pm 0.05     $ &  58  \\
                          & 0.68617   & $16.49 \pm 0.04     $ &  32  \\
LBQS-1435-0134            & 0.26228   & $14.96 \pm 0.02     $ &  25  \\
                          & 0.29907   & $15.30 \pm 0.02     $ &  25  \\
                          & 0.39214   & $15.03 \pm 0.02     $ &  25  \\
                          & 0.43834   & $14.76 \pm 0.02     $ &  25  \\
                          & 0.61283   & $15.30 \pm 0.02     $ &  25  \\
                          & 0.68124   & $15.54 \pm 0.02     $ &  25  \\
PG-1522+101               & 0.51841   & $16.32 \pm 0.2      $ &  16  \\
                          & 0.57179   & $15.31 \pm 0.06     $ &  17  \\
                          & 0.67518   & $15.87 \pm 0.01     $ &  25  \\
                          & 0.72879   & $16.60 \pm 0.09     $ &  26  \\
Q0232-042                 & 0.32239   & $16.14 \pm 0.03     $ &  30  \\
                          & 0.73888   & $16.64 \pm 0.08     $ &  35  \\
                          & 0.80773   & $15.60 \pm 0.05     $ &  32  \\
PG1630+377                & 0.27395   & $16.92 \pm 0.04     $ &  44  \\
                          & 0.27821   & $14.72 \pm 0.02     $ &  25  \\
                          & 0.41774   & $15.72 \pm 0.02     $ &  25  \\
                          & 0.41856   & $14.67 \pm 0.04     $ &  25  \\
                          & 0.8111    & $15.52 \pm 0.03     $ &  25  \\
                          & 0.91449   & $15.81 \pm 0.01     $ &  25  
\enddata

\tablenotetext{a}{In these 71 AGN sight lines we find 7 Lyman Limit systems
($\log N_{\rm HI} \geq 17.2$) and 214 partial Limit limit absorbers ($15.0 \leq \log N_{\rm HI} < 17.2$).  
We list their redshifts, \HI\ column densities and lower-limit statistical uncertainties, and 
doppler parameters ($b$) derived by fitting \Lya\ and higher Lyman series absorbers.
When the doppler parameter is unconstrained, we used mean $b = 25$~\kms\
to propagate uncertainties in the continuum slope of individual spectra.}  

\end{deluxetable}

 \end{document}